\documentclass[aps,prc,twocolumn,showpacs,preprintnumbers,superscriptaddress,
float,longbibliography]{revtex4-1}

\usepackage{color}
\usepackage[dvipsnames]{xcolor}
\usepackage{graphicx}
\usepackage{dcolumn}
\usepackage{bm}
\usepackage{appendix}
\usepackage{multirow}
\usepackage[colorlinks=true, pdfstartview=FitV, linkcolor=blue,
citecolor=blue, urlcolor=blue]{hyperref}
\usepackage{amsmath}
\usepackage{mathrsfs}

\usepackage{xcolor}
\usepackage[normalem]{ulem}

\def\be{\begin{equation}}
\def\ee{\end{equation}}
\def\bea{\begin{eqnarray}}
\def\eea{\end{eqnarray}}

\def\kompost{K\o{}MP\o{}ST }

\renewcommand{\vec}{\bm}

\def\nB{f_{_{\rm B}}}
\def\pT{p_{_{\rm T}}}
\newcommand{\sqrts}{\sqrt{s_\mathrm{NN}}}

\begin{document}


\title{Thermal Dilepton Polarization and Dynamics of the QCD Plasma in Relativistic Heavy-Ion Collisions}

\author{Xiang-Yu Wu}
\affiliation{%
    Department of Physics, McGill University, 
    3600 University Street, Montreal, Quebec H3A 2T8, Canada}
\author{Han Gao}
 \affiliation{%
   Department of Physics, McGill University, 
   3600 University Street, Montreal, Quebec H3A 2T8, Canada}
\author{Bailey Forster}
\affiliation{%
    Department of Physics, McGill University, 
    3600 University Street, Montreal, Quebec H3A 2T8, Canada}
\author{Charles Gale}
 \affiliation{%
   Department of Physics, McGill University, 
   3600 University Street, Montreal, Quebec H3A 2T8, Canada}
\author{Greg Jackson}
\affiliation{%
   SUBATECH, 
   Universit\'e de Nantes, IMT Atlantique, IN2P3/CNRS,
4 rue Alfred Kastler, La Chantrerie BP 20722, 44307 Nantes, France}
\author{Sangyong Jeon}
 \affiliation{%
   Department of Physics, McGill University, 
   3600 University Street, Montreal, QC H3A 2T8, Canada}
\date{\today}

\begin{abstract}
We present the first theoretical study of the polarization of lepton pairs produced in $\sqrts = 5.02 $ TeV Pb+Pb collisions at the LHC, 
using next-to-leading order (NLO) dilepton emission rates. 
These calculations employ a multistage framework to simulate the 
evolution of relativistic heavy-ion collisions, and to explore 
the sensitivity of polarization to early times. 
It is found that the intermediate invariant-mass dileptons are indeed probes of the thermal equilibration process, and 
go beyond the reach of hadronic observables. We compute the polarization anisotropy coefficient obtained with LO dilepton rates, and show that the LO and NLO results differ radically, 
both in trend and in magnitude,
at low and intermediate lepton pair invariant masses. 
\end{abstract}

\maketitle

{\it \textbf{Introduction.}---}
Relativistic heavy-ion collisions 
produce the quark-gluon plasma (QGP), an exotic state of matter described  by  quantum chromodynamics~\cite{Jacak:2012dx}. 
As the QGP expands and cools over the duration of a collision 
between two nuclei, the quarks can emit electromagnetic (EM) 
radiation which escapes the strongly interacting medium~\cite{Gale:2021emg,Paquet:2015lta,vanHees:2011vb,Gale:2003iz,Gale:1987ki,vanHees:2007th}. 
Therefore, photon and dilepton measurements 
have the potential to  provide direct information about the hot medium, 
such as electric conductivity~\cite{Rapp:2024grb}, chemical equilibrium~\cite{Wu:2024pba,Garcia-Montero:2024lbl,Coquet:2021lca}, 
magnetic fields~\cite{Wang:2020dsr,Wang:2021ebh,Wang:2022jxx,Kimura:2024gao}, 
and most notably, early-stage 
temperature~\cite{NA60:2008ctj,Churchill:2023zkk,Churchill:2023vpt}. Experimentally, 
EM probes have been recently measured at the Large Hadron Collider (LHC)~\cite{ALICE:2023jef} 
and the Relativistic Heavy-Ion Collider 
(RHIC)~\cite{STAR:2023wta,STAR:2015tnn,STAR:2024bpc}.
\par
The $\pT$ spectrum of direct photons is sensitive to local 
hydrodynamical 
flow, and can therefore inform the theoretical modeling~\cite{Gale:2021emg}.  In contrast, the invariant mass spectrum of dileptons~---~a complementary probe~---~is impervious to flow effects. 
For invariant mass $M\gtrsim 1$~GeV, 
an 
important source
of dileptons is   
quark-antiquark 
annihilation into a virtual photon (labeled from here on LO, a leading-order contribution)~\cite{NA60:2008dcb,STAR:2024bpc,Rapp:2013nxa}~
\footnote{%
  This channel is also included in the Drell-Yan process 
  where initial quarks are sourced from the nPDFs of the colliding ions.
}, 
yielding 
thermal, and also preequilibrium dileptons~\cite{Coquet:2021lca,Wu:2024pba,Garcia-Montero:2024lbl}.  
However, 
corrections from the strong interaction occur in 
the intermediate mass range (IMR), $1~{\rm GeV} \lesssim M \lesssim 3$~GeV, 
and even become dominant in the low mass range (LMR): $M \lesssim 1$~GeV. 
For completeness, one also defines a high mass range (HMR) as $M \gtrsim 3$~GeV. 
We will focus on the thermal 
mechanism  
in what follows.

Because of the gluon's abundance in a hot QGP medium, 
new channels open up at next-to-leading order (NLO) 
which includes: Compton scattering 
($g\,q\to \gamma^* \, q$ and $g\,\bar q\to \gamma^* \, \bar q$), 
as well as modified annihilation 
($q \, \bar q \to \gamma^* \, g$ and $q \, \bar q \, g \to \gamma^*$, 
with the latter being kinematically suppressed for $M\simeq 0$). 
Note that the above ``real" gluon emissions need to be combined 
with the ``virtual'' (1-loop) corrections to $q\,\bar q \to \gamma^*$ 
to obtain a finite result at strict NLO~\cite{Laine:2013vma,Jackson:2019mop}. 
As $M\to 0$, the strict expansion becomes invalid due to 
the physical necessity of thermal screening and 
the Landau-Pomeranchuck-Migdal (LPM) effect. 
These are addressed in an effective theory (for small $M$) 
which resums naively higher-order terms~\cite{Arnold:2001ba,Arnold:2001ms,Aurenche:2002pc,Aurenche:2002wq}.
Both regimes can be systematically combined in the spectral function 
obtained from thermal field theory, to cover both small and 
large $M$~\cite{Ghisoiu:2014mha,Jackson:2019yao,Ghiglieri:2021vcq}. 
Together, the NLO and LPM contributions~\footnote{%
From this point on, we consider the ``NLO contributions'' to include the LPM resummation for small $M$ as in Refs.~\cite{Ghisoiu:2014mha,Jackson:2019yao}.} 
are found to qualitatively enhance 
the observed thermal dilepton spectrum in the 
LMR, 
as well as in the 
IMR~\cite{Churchill:2023zkk, Churchill:2023vpt,Churchill:2023hog}. 
\par
In addition to mass and $\pT$ spectra, 
the polarization of EM radiation provides another unique tool 
to study medium properties. For example, 
the angular distribution of the thermal lepton pair in 
the rest frame of the virtual 
photon~\cite{Speranza:2018osi,Wei:2024lah,Seck:2023oyt}~\footnote{%
  Also the centre-of-mass frame of the dilepton.
} 
is expected to be sensitive to plasma anisotropy~\cite{Coquet:2023wjk}; the same can be said for the polarization of real photons~\cite{Hauksson:2023dwh}. 
\par
This work aims to provide a realistic study of 
the polarization of thermal dileptons produced 
in heavy-ion collisions 
at the LHC, using 
state-of-the-art 
multistage modeling calibrated 
to reproduce hadronic observables at 
$\sqrts =5.02$~TeV~\cite{Wu:2024pba} 
together with 
strong coupling corrected EM production rates.
We introduce the polarization coefficient,  
and highlight its explicit dependence on the local flow and the thermal spectral densities. 
We show how dilepton polarization in the IMR has the potential 
to reveal the importance of NLO contributions, and thus  
the role of gluons in the medium. 
We also show results for the LMR, 
with the understanding that processes involving 
composite hadrons will play a crucial role there~\cite{Rapp:2013nxa}, 
while being subdominant in the IMR. 
We also estimate 
polarization results for the elusive 
preequilibrium phase of relativistic heavy-ion collisions.

{\it \textbf{Theoretical Setup.}---}
We consider a fluid cell in thermodynamic equilibrium, 
characterized by a four-velocity $u^\mu$ and a temperature $T$~\footnote{%
  The temperature is defined in the fluid rest frame.
}.
The differential emission rate $R_{\ell \bar \ell}$ of dileptons 
is~\cite{Kapusta:2023eix}
\begin{equation}\label{eq:d6rate}
    E_+ E_- \frac{{\rm d} R_{\ell \bar \ell}}{{\rm d}^3 \vec p_+ {\rm d}^3 \vec p_-} 
    \; =\; 
    -\frac{2e^4}{(2\pi)^6} 
    \frac{\sum_i^{N_{\rm f}} Q_i^2}{K^4} \,
    L^{\mu\nu} \,
    \rho_{\mu\nu} \,
    \nB(\omega)
    \; .
\end{equation}
Here, 
$K = P_+ + P_-$ is the four-momentum of the virtual photon, 
$\nB(\omega) \equiv \frac{1}{e^{\omega/T} -1}$ is the Bose-Einstein distribution, 
and $e Q_i$ is the charge of quark flavor $i\,$. In this work $N_{\rm f} = 3$. 
The leptonic tensor is $L^{\mu\nu} \equiv 
P_+^\mu P_-^\nu + P_-^\mu P_+^\nu - g^{\mu\nu} (P_+ \cdot P_- + m_\ell^2)\,$,
and the photon spectral function $\rho_{\mu\nu}$ is 
a function of 
the two invariants
$\omega \equiv u\cdot K$ and $k \equiv \sqrt{\omega^2 - K^2}\,$ (which correspond to the energy and momentum of the virtual photon in the medium's rest frame), formally obtained by analytic continuation of 
the Euclidean correlator $G_{\mu\nu}$\,:
\begin{eqnarray}
  \rho_{\mu\nu}(\omega,\vec k)
  &\equiv&
  - \, {\rm Im} \bigg[ \int_0^{1/T} {\rm d} \tau 
  \, e^{i \omega_n \tau } 
  G_{\mu\nu}(\tau, \vec k )
  \bigg]_{i\omega_n \to \omega + i0^+} \, ,\nonumber\\
  G_{\mu\nu}(\tau, \vec k) 
  \; &=& \;
  \int {\rm d}^3 \vec x \, 
  e^{ - i \vec k \cdot \vec x} 
  \big\langle 
  J_\mu(\tau, \vec x) J_\nu(0,\vec 0)
  \big\rangle_T \, ,
  \label{eq:rho_def}
\end{eqnarray}
where $J_\mu = \bar \psi \gamma_\mu \psi$ is the EM current, 
$\langle ... \rangle_T$ denotes the thermal average, 
and $\tau$ is the imaginary-time~\cite{Laine:2016hma}.
\par 
As the medium velocity $u^\mu$ specifies a 
preferred 
direction, 
we can decompose $\rho^{\mu\nu}$ into longitudinal and transverse parts, respectively as~\cite{Weldon:1982aq}
\begin{equation}\label{eq:projectors}
    \rho_{_{\rm L}} 
    \; \equiv \; 
    -\frac{K^2 \; \rho_{\mu\nu}u^\mu u^\nu}{(u\cdot K)^2 - K^2}
    \,,\quad 
    \rho_{_{\rm T}} 
    \; \equiv \;
    \frac{\rho^\mu_\mu - \rho_{_{\rm L}}}2 
    \; .
\end{equation}
It is also useful to define the ``vector channel'' and the polarization difference, namely
\begin{equation}
    \rho_{_{\rm V}} 
    \; \equiv \; 
    \rho^\mu_\mu = \rho_{_{\rm L}} + 2\rho_{_{\rm T}}
    \,,\quad 
    \rho_{_\Delta} 
    \; \equiv \;
    \rho_{_{\rm T}} - \rho_{_{\rm L}} = \frac{\rho_{_{\rm V}} - 3\rho_{_{\rm L}}}2.
\end{equation}
\par 
From Eq.~(\ref{eq:d6rate}),  one can integrate the relative 
momentum of the $\ell \bar \ell$ pair to obtain the dilepton production rate 
in terms of the four momentum of the virtual photon 
alone, 
\begin{equation}\label{eq:drdk}
    \frac{{\rm d}R_{\ell\bar\ell}}{{\rm d}^4K} 
    \; = \;  
    \frac{2 \alpha_{\rm em}^2}{9\pi^2} 
    \ \frac{ \sum_i^{N_{\rm f}} Q_i^2}{K^2}
    \, B\bigg(\frac{m_\ell^2}{K^2}\bigg) \,
    \rho_{_{\rm V}} 
    \, \nB(\omega)
    \; ,
\end{equation}
where $B(\xi) \equiv (1+2\xi)\sqrt{1-4\xi}$ is a phase-space factor. 
As the equilibrium rate in Eq.~\eqref{eq:drdk} is proportional to $\rho_{_{\rm V}}$, 
 observables such as dilepton invariant mass spectra 
and elliptic flow cannot differentiate between 
$\rho_{_{\rm T}}$ and $\rho_{_{\rm L}}$.
\par
Nonperturbative constraints on the spectral 
function can be obtained from lattice QCD simulations, 
where 
$G_{\mu\nu}(\tau,\vec k)$
is measured at 
fixed momenta $k= 2\pi \, n/(a N_s)$ where $n$ is an integer, 
$a$ is the lattice spacing and $N_s$ is the number of spatial sites. 
In practice, the analytic continuation to {\em real} frequencies is fraught with difficulty~\cite{Meyer:2011gj}. 
Yet significant progress has been made, first focusing on 
$\rho_{_{\rm V}}$~\cite{Ghiglieri:2016tvj}. 
However, the vector channel is insensitive to thermal physics 
owing to the large ultraviolet vacuum component of the photon self-energy. 
For this reason, it was suggested to consider the correlator $\rho_\Delta$ which vanishes in vacuum and is highly suppressed for large $\omega$~\cite{Brandt:2017vgl,Ce:2022fot,Meyer:2023ntn}.
Recently, Ref.~\cite{Ali:2024xae} of the HotQCD Collaboration~\cite{HotQCD:2014kol} obtained estimates for $\rho_{_\Delta}(\omega,k)$ for $2$+$1$ flavors~\footnote{%
Although these results were at a pion mass $m_\pi \simeq 320$~MeV and not continuum extrapolated, the lattice was quite large with $N_s = 96$ and $a \simeq 7$~GeV$^{-1}$.} at $T \simeq 220$~MeV with spatial momenta 
ranging from $k  \simeq 0.5$ to $1.4$~GeV. 
Ubiquitously, $\rho_{_\Delta}$ decreases as a function of $\omega\,$, from positive at $\omega = k$ and then quickly becoming negative for $\omega > k\,$.
This is consistent with perturbation theory, where 
both $\rho_{_{\rm T}}$ and $\rho_{_{\rm L}}$ 
have been worked out at strict NLO~\cite{Laine:2013vma,Jackson:2019mop}, 
and also generalized to finite baryon density~\cite{Jackson:2022fqj,Churchill:2023vpt}. 
\par 
Complementary to the lattice agenda, $\rho_{_\Delta}$ is also relevant for heavy-ion experiments, 
when it comes to
the angular distribution of the final $\ell \bar \ell$ pair.
The latter observable, in the rest frame of the virtual photon, 
discriminates between the longitudinal and transverse polarization. 
This angular distribution can be parametrized as 
\begin{eqnarray}\label{eq:defpol}
   \frac{{\rm d}N}{{\rm d}^4 K \, {\rm d}\Omega_\ell } 
   &\propto & 
   1 + \lambda_\theta \cos^2 \theta_\ell + \lambda_\phi \sin^2\theta_\ell \cos 2\phi_\ell 
   \nonumber \\[-1mm]
   &+&  \lambda_{\theta\phi} \sin 2\theta_\ell \cos \phi_\ell + \lambda_\phi^\perp \sin^2\theta_\ell \sin 2\phi_\ell 
   \nonumber \\[1mm]
   &+& \lambda^\perp_{\theta \phi} \sin2\theta_\ell \sin \phi_\ell 
   \; .
\end{eqnarray}
A common choice of the $\gamma^*$ rest frame is the so-called 
helicity (HX) frame where the $z$ axis aligns with the momentum 
of the virtual photon~\cite{Faccioli:2010kd}. 
Here, polar angles $\Omega_{\ell}=(\theta_\ell,\phi_\ell)$ 
are defined in a way such that $\ell^+$ in the HX frame has the 
three-momentum 
$\vec{l}_+ = l \, (\sin\theta_\ell \cos \phi_\ell,\sin\theta_\ell \sin \phi_\ell,\cos\theta_\ell)$. 
By performing a change of variables and integrating out the radial part of dilepton 
relative momentum in 
Eq.~\eqref{eq:d6rate}, we arrive at an expression 
for the 
polarization coefficient $\lambda_\theta$ 
of a single fluid cell 
at temperature $T$\,: 
\begin{equation}\label{eq:lambda_theta}
    \lambda_{\theta} 
    \; =\; 
    \frac{ 
    3
    ( \chi - \frac{1}{3})
    \left( 1- 4 \xi\right)\rho_{_\Delta}
    }{
    \frac{4}{3}
    ( 1+ 2 \xi) \rho_{_{\rm V}} 
    -
    (\chi - \frac{1}{3})
    ( 1-4 \xi) \rho_{_\Delta}
    }
    \; , 
\end{equation}
where $\xi \equiv \frac{m_\ell^2}{K^2}$
and 
$\chi \equiv \frac{(\vec u_* \cdot \hat k)^2}{\vec u^2_*}\,$
with $\vec u_*$ denoting the fluid velocity viewed in the HX frame. 
It is related to $\vec u$ in the lab frame by a Lorentz boost
\begin{equation} \label{eq:ustar}
    \vec u_* = \vec u + \left( \frac \omega M - 1\right) (\vec u \cdot \hat k)\hat k - \frac{u_0 \vec k}{M}.
\end{equation}
One can work out the other coefficients from Eq.~\eqref{eq:defpol}, 
in a similar manner. Nevertheless, 
$\lambda_\theta$ is the only coefficient that does not vanish 
for a medium at rest, making it an  indicator of $\rho_{_\Delta}$ 
that is less sensitive to local flow conditions. 
In this Letter, we focus on $\lambda_\theta$. 
\par 
\begin{figure}[t!]
	\centering
	\includegraphics[width=\linewidth]{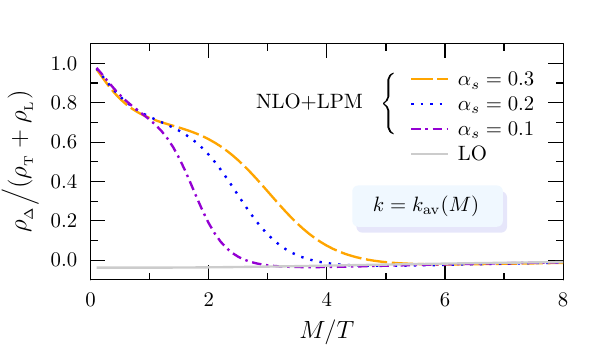}
    \vspace{-6mm}
    \caption{
    Approximation to $\lambda_\theta$ from the ratio of spectral functions in Eq.~\eqref{eq:lambda_approx}, as a function of the invariant mass $M$ with $k$ fixed by Eq.~\eqref{eq:k2av}. The resummed NLO result is shown for fixed coupling, $\alpha_s = \{0.1,0.2,0.3\}$. 
    }
	\label{fig:lambda_k2av}
\end{figure}
We note 
that Eq.~\eqref{eq:lambda_theta} is 
closely connected with $\rho_{_\Delta}$~\cite{Seck:2023oyt}.
Indeed, for a plasma at rest $u^\mu=(1,\vec 0)$~(i.e. $\chi = 1$), 
evaluating Eqs.~\eqref{eq:lambda_theta} and \eqref{eq:ustar} 
and assuming $m_\ell \approx 0$ 
yields
\begin{equation} \label{eq:lambda_approx}
    \lambda_\theta 
    \; =\; 
    \frac{
    (1-4\xi)
    \big( \rho_{_{\rm T}} - \rho_{_{\rm L}} \big)
    }{
    \rho_{_{\rm L}} + 
    (1 + 4 \xi)
    \rho_{_{\rm T}}
    } 
    \; \simeq \; 
    \frac{
    \rho_{_\Delta}
    }{ \rho_{_{\rm T}} + \rho_{_{\rm L}} }
    \; ,
\end{equation}
which is displayed in Fig.~\ref{fig:lambda_k2av}.

Before embedding the fully differential rates on a hydrodynamical background, 
it is instructive to analyze Eq.~\eqref{eq:lambda_approx} 
for small and large invariant masses, omitting flow conditions.
Because the spectral functions are controlled by $M/T$, 
the large $M$ limit can be viewed as $T\to 0$ for a fixed $M$. 
The thermal medium becomes irrelevant from the 
perspective 
of a highly virtual photon. 
In this case, $\gamma^*$  
behaves as it was a massive vector field 
randomly produced in vacuum, 
having no preference in its orientation. 
Hence, $\lambda_\theta \to 0$ for both LO and NLO in the large $M$ limit, as we shall see below. 
This asymptotic behavior is indeed confirmed more quantitatively by the operator product expansion~\cite{Caron-Huot:2009ypo,Laine:2010tc}, where it was found that 
$\rho_{_{\Delta}}\sim  k^2 (T/M)^4$ while $\rho_{_{\rm T,L}}\sim M^2$ for $M\gg T,k$.
For the opposite limit $M/T\to 0$, the virtual photon is more akin to a real photon, lacking the longitudinal polarization, so  that $\lambda_\theta \simeq (\rho_{_{\rm T}}-\rho_{_{\rm L}})/(\rho_{_{\rm T}} + \rho_{_{\rm L}}) \to 1$. 
This is indeed observed at NLO, as revealed by Fig.~\ref{fig:lambda_k2av} where, for illustration, 
we evaluate the spectral functions at $k = k_{\rm av}(M)$ and $\omega = \sqrt{M^2 + k^2}$, with 
\begin{equation} \label{eq:k2av}
   k_{\rm av}^2 (M) \; \equiv \; 
   \frac{
   \int_0^\infty {\rm d}k \, k^4 \exp(-\frac{\sqrt{M^2+k^2}}{T})
   }{
   \int_0^\infty {\rm d}k \, k^2 \exp(-\frac{\sqrt{M^2+k^2}}{T})
   }
   \; = \; 
   \frac{ 3 M T \, K_3(\frac{M}{T}) }{ K_2(\frac{M}{T}) }
   \; , 
\end{equation}
$K_n$ being the modified Bessel function of the second kind.
From Fig.~\ref{fig:lambda_k2av}, we note that the limit for $M \ll T$ is robust even when varying $\alpha_s\,$. 
This physics is absent from LO calculations as the process $q\bar q\to \gamma$ is kinematically forbidden, making the LO incapable of converging to the real-photon limit, owing to the lack of channels containing gluons and created at NLO. 
Therefore the 
NLO correction is essential for studying polarization, 
especially in LMR, 
but also in the IMR {{\footnote{Note that calculations to NNLO and beyond need to respect the fact that $\rho_{\rm L}$ = 0 for $M \to 0$, and $\rho_{\rm T} = \rho_{\rm L}$ for $M \to \infty$.}}}The rest of our study will use $\alpha_s = 0.3$, 
a value which is consistent with the bulk of 
phenomenological studies performed for collisions 
performed at the LHC~\cite{Arslandok:2023utm}.

{\it \textbf{Thermal Dilepton Phenomenology.}---}
The relativistic event-by-event heavy-ion collisions of Pb+Pb at $\sqrts = 5.02$ TeV are simulated using the 
iEBE-MUSIC framework~\cite{Schenke:2010nt,Schenke:2010rr,Paquet:2015lta}, 
adopting the same model setup as in Ref.~\cite{Wu:2024pba}. The production of lepton pairs is calculated as it was in Ref. \cite{Wu:2024pba}. 
The local temperature $T(X)$,
and local flow velocity $u^\mu(X)$ for each fluid cell at space-time point $X=(t,\vec{x})$ are obtained to yield the polarization observables (at LHC energies, the baryon chemical potential $\mu_B$ 
can be neglected). 
For a given lab-frame virtual photon four-momentum $P$,  
the final $\lambda_\theta$ is a weighted average over all fluid cells, namely
\begin{equation}\label{eq:average_lambda}
    \lambda_\theta(P)
    \; = \;  
    \frac{
    \int {\rm d}^4X \,
    \lambda_{\theta}\big(P;T(X)
    \big) {\cal R}\big(P;T(X)
    \big)
    }{
    \int {\rm d}^4 X\, 
    {\cal R}\big(P;T(X)
    \big)
    },
\end{equation}
where 
$
{\cal R}\big(P;T
\big) 
\equiv 
\left[\frac{{\rm d}R_{\ell \bar \ell}}{{\rm d}^4K}\right]
\big(K;T
\big)\big/(1 + \lambda_\theta\big(P,T
)/3\big)
$ 
is the weight function obtained from integrating over 
${\rm d}\Omega_\ell$ of the both sides of Eq.~(\ref{eq:defpol}), 
and $K^\mu = \Lambda^\mu_\nu(X) P^\nu$ is the virtual photon four-momentum 
in the local rest frame of the fluid cell at $X$, 
boosted according to the local $u^\mu(X)\,$. 
The invariant mass and transverse momentum, 
dependent 
$\lambda_\theta(M)$ and $\lambda_\theta (\pT)$, 
are obtained in a similar way with complementary kinematics integrated out.

\begin{figure}[hb]
	\centering
	\includegraphics[width=1.0\linewidth]
    {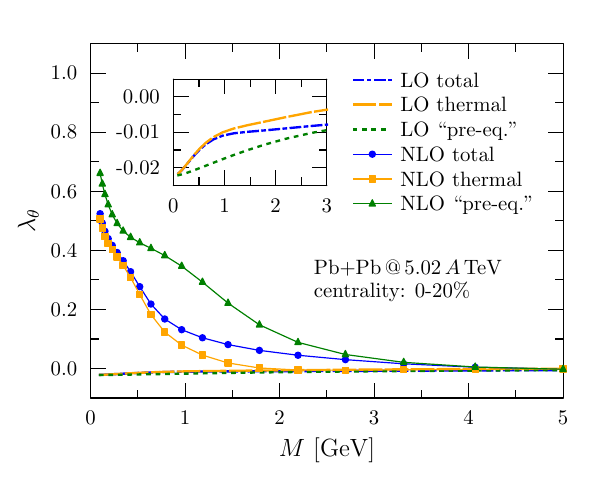}
\vspace{-6mm}
 \caption{
 The anisotropy coefficient $ \lambda_\theta $ 
 of dilepton polarization with LO and NLO, induced by  various sources, as a function of invariant mass $M$ 
 in the 
 0\%-20\% centrality at $\sqrts = 5.02$~TeV  Pb+Pb collisions at the LHC. The inset figure shows the LO results scaled for clarity. 
 Here, and in the rest of this Letter, dilepton results are for dielectrons.
 }
	\label{fig:labth_M}
\end{figure}

Figure~\ref{fig:labth_M} 
shows the anisotropy coefficient $\lambda_\theta $ 
as a function of the invariant mass $ M $ in the 0\%-20\% centrality class
for 
Pb+Pb collisions at $\sqrts = 5.02$ TeV. 
Notably, in the LMR 
the inclusion of NLO corrections changes the anisotropy coefficient qualitatively: 
$ \lambda_\theta(M)$ shifts from a near-zero negative to a sizable positive value after considering the NLO correction (this feature is also evident in Fig.~\ref{fig:lambda_k2av}). 
Physically, this indicates that $q\bar q$ annihilation tends to produce longitudinal polarization in the HX frame, 
while the gluon Compton scatterings are
more likely to result in $\gamma^*$ with transverse polarization. 
\par
We note that $\lambda_{\theta}$ 
given in Eq.~\eqref{eq:lambda_theta}, assuming $m_\ell = 0$, 
satisfies 
$
- \frac13 \leq \lambda_\theta|_{M\to 0}
= \frac{1-3\chi}{\chi-3}
\leq 1\,
$. 
The upper bound is saturated for $\chi =1$ (i.e. $\vec u_* = 0$), 
as consistent with the behavior of 
$\rho_{_\Delta}/(\rho_{_{\rm T}} + \rho_{_{\rm L}})$ in Fig.~\ref{fig:lambda_k2av}.  
The lower bound occurs  
when $\chi=0$ (i.e. $\vec{u}_* \cdot \hat k = 0$), 
giving 
$\lambda_\theta =\frac{- \rho_{_\Delta}}{\rho_{_{\rm L}} + 3 \rho_{_{\rm T}}}\to -\frac13 $ for 
$M=0\,$. 
Therefore, once local flow conditions are included, 
$\lambda_\theta$ at $M \simeq 0$ limit will be slightly less 
than 1, as realized in Fig.~\ref{fig:labth_M} for the NLO result. 
Nevertheless, although diluted by the plasma kinematics, the results obtained with NLO and LO rates are still radically different in the limit $M\to0$, since the LO spectral function fails to capture the relevant physics in the IMR.  
In the HMR, 
QGP dileptons become almost unpolarized as $\lambda_\theta \approx 0$, consistent again with the pure thermal prediction in Fig.~\ref{fig:labth_M}. 
However, the Drell-Yan process, 
the consideration of which lies beyond the scope of this study, 
is the main source of dilepton production and polarization in the HMR. 
{Figure~2 also shows an estimation of the effect of a preequilibrium phase on the polarization coefficient. Here, the preequilibrium dynamics are addressed using \kompost with the same setup as in Ref.~[2], namely evaluating thermal rates using an extracted effective temperature.} 
 That prehydro addition, included in the ``total'' polarization,  has a noticeable but relatively modest effect here.  

\begin{figure}[t]
	\centering
    \includegraphics[width=1.0\linewidth]{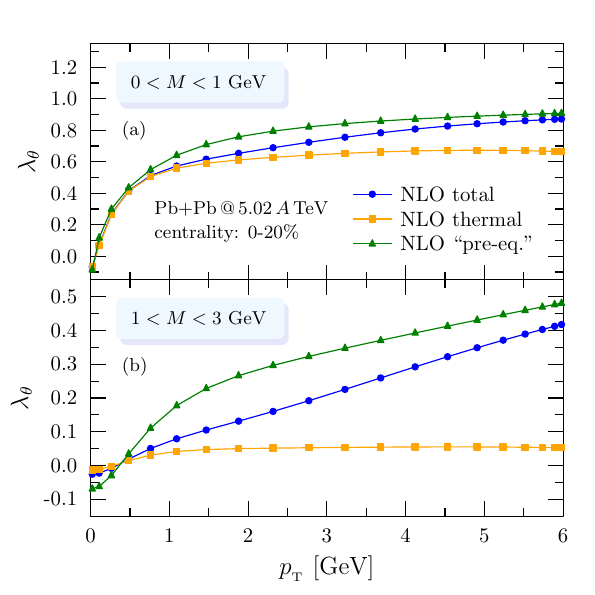}
    \vspace{-6mm}
     \caption{
     The anisotropy coefficient $ \lambda_\theta $ of dilepton polarization evaluated at NLO,  induced by  various sources, as a function of transverse momentum $\pT$ in the LMR [panel (a)] and IMR [panel (b)], for the 0\%-20\% centrality class in  Pb + Pb collisions at $\sqrts = 5.02$~TeV.
     }
	\label{fig:labth_pT0}
\end{figure}

In Fig.~\ref{fig:labth_pT0}, we show $\lambda_{\theta}$ 
as a function of transverse momentum $\pT$ in the LMR and the IMR, 
where contributions from different stages of the fireball evolution are also displayed. {Our conclusions apply to the IMR, where the hadronic stage is known to have limited influence~\cite{vanHees:2007ma,Vujanovic:2013jpa}, but the LMR is also displayed for completeness. }{Note that it is known that  open charm and beauty meson decays constitute a major background in IMR dilepton measurements. However, planned next-generation detectors will have
vertex detection capabilities, rendering the identification of semileptonic decays more efficient and enabling a subtraction of that contribution. Note that the heavy flavor contribution to dilepton polarization is
complicated by the fact that, for example, the production of $D \bar{D}$ (followed by semileptonic
decays) does not involve a ``clean'' intermediate state of just a virtual photon,
but rather has a combinatorial origin. Such studies are well suited for event generators.}

In a given range of invariant mass,  
$\lambda_\theta$ increases as a function of $\pT$. 
The behavior is consistent with that observed earlier: 
as the transverse momentum grows, the influence of the invariant mass shrinks, bringing the kinematics closer to those of real photons. 
In that limit, the transverse polarization dominates over the longitudinal. 
In the fluid dynamical environment, the exact value is also affected by the averaging over the flow. 
For the opposite limit $\pT \to 0 $, we observe that $ \lambda_{\theta} $ does not vanish,
which 
can also be attributed to the nonzero flow velocity of the medium in the HX frame of the dilepton.

Since  
gluon-mediated processes only contribute to dilepton polarization 
at NLO (and beyond),  
the corresponding enhancement of $\lambda_\theta$ in the IMR window, 
if experimentally confirmed, 
could be taken as a flag for 
such elementary QCD interactions.
Moreover, the behavior of 
$\lambda_\theta(\pT)$ 
could indicate the 
relative abundance of quarks and gluons: 
a larger $\lambda_\theta(\pT)$ means a greater gluon-quark ratio during the preequilibrium stage than that in thermal equilibrium, as is clear in the bottom panel of Fig.~\ref{fig:labth_pT0}. 
It is therefore very promising that the $\pT$ dependence of $\lambda_{\theta}$ 
in the IMR will offer valuable insights into the preequilibrium 
stage of heavy-ion collisions in future studies.

{\it \textbf{Summary.}---}
In this Letter, 
we perform the first study of dilepton polarization phenomenology 
using 
multistage heavy-ion collision simulations 
for $\sqrts = 5.02$~TeV Pb+Pb collisions at the LHC. 
We employed NLO thermal emission rates in our analysis, 
which were coupled with the local temperature and flow velocity 
provided by the complete evolution history of the heavy-ion simulation.

We observe 
an overall sign change and a significant enhancement in $\lambda_\theta$ when going from LO to NLO. 
This better matches the physical expectation 
that at small invariant mass, the virtual photon behaves like a real photon with {\em two} transverse polarization modes. 
In contrast, the LO calculation does not capture this aspect due to the vanishing phase space for $M\to 0\,$, where the NLO result is actually the ``leading'' contribution. 
At large invariant masses, the NLO result is a small correction to the LO one and both reflect the lack of a preferred polarization direction for $M \gg T$.
Underlying this dramatic difference between orders in the perturbative calculation, is the participation of (thermal) gluons in the production reactions. 
The importance of gluons is readily supported from nonperturbative lattice studies, where the current-current correlators do not change qualitatively when going from quenched to $2$+$1$ flavor QCD~\cite{Brandt:2017vgl,Ce:2022fot,Ali:2024xae}.

We also explore the dilepton polarization phenomenon induced during 
the preequilibrium stage and the QGP phase. 
It is found that dilepton polarization can be highly sensitive to the 
preequilibrium evolution; 
we find that the effects of the preequilibrium phase on 
dilepton polarization become more pronounced when considering 
the transverse momentum $\pT$-dependence in the IMR. 
Therefore, studies of the preequilibrium stages in heavy-ion collisions will benefit from those of dilepton polarization. 

{\it \textbf{Acknowledgements.}---}
This work was supported in part by the Natural Sciences and Engineering Research Council of Canada  (NSERC) [SAPIN-2020-00048 and SAPIN-2024-00026], 
and in part by US National Science Foundation (NSF) under grant number OAC-2004571. 
G.J. was funded by the Agence Nationale de la Recherche (France), under grant ANR-22-CE31-0018 (AUTOTHERM). 
Computations were made on the B\'eluga super-computer system from McGill University, managed by Calcul Qu\'ebec and by the Digital Research Alliance of Canada.

\bibliography{references}

\begin{thebibliography}{64}%
\makeatletter
\providecommand \@ifxundefined [1]{%
 \@ifx{#1\undefined}
}%
\providecommand \@ifnum [1]{%
 \ifnum #1\expandafter \@firstoftwo
 \else \expandafter \@secondoftwo
 \fi
}%
\providecommand \@ifx [1]{%
 \ifx #1\expandafter \@firstoftwo
 \else \expandafter \@secondoftwo
 \fi
}%
\providecommand \natexlab [1]{#1}%
\providecommand \enquote  [1]{``#1''}%
\providecommand \bibnamefont  [1]{#1}%
\providecommand \bibfnamefont [1]{#1}%
\providecommand \citenamefont [1]{#1}%
\providecommand \href@noop [0]{\@secondoftwo}%
\providecommand \href [0]{\begingroup \@sanitize@url \@href}%
\providecommand \@href[1]{\@@startlink{#1}\@@href}%
\providecommand \@@href[1]{\endgroup#1\@@endlink}%
\providecommand \@sanitize@url [0]{\catcode `\\12\catcode `\$12\catcode
  `\&12\catcode `\#12\catcode `\^12\catcode `\_12\catcode `\%12\relax}%
\providecommand \@@startlink[1]{}%
\providecommand \@@endlink[0]{}%
\providecommand \url  [0]{\begingroup\@sanitize@url \@url }%
\providecommand \@url [1]{\endgroup\@href {#1}{\urlprefix }}%
\providecommand \urlprefix  [0]{URL }%
\providecommand \Eprint [0]{\href }%
\providecommand \doibase [0]{http://dx.doi.org/}%
\providecommand \selectlanguage [0]{\@gobble}%
\providecommand \bibinfo  [0]{\@secondoftwo}%
\providecommand \bibfield  [0]{\@secondoftwo}%
\providecommand \translation [1]{[#1]}%
\providecommand \BibitemOpen [0]{}%
\providecommand \bibitemStop [0]{}%
\providecommand \bibitemNoStop [0]{.\EOS\space}%
\providecommand \EOS [0]{\spacefactor3000\relax}%
\providecommand \BibitemShut  [1]{\csname bibitem#1\endcsname}%
\let\auto@bib@innerbib\@empty
\bibitem [{\citenamefont {Jacak}\ and\ \citenamefont
  {Muller}(2012)}]{Jacak:2012dx}%
  \BibitemOpen
  \bibfield  {author} {\bibinfo {author} {\bibfnamefont {Barbara~V.}\
  \bibnamefont {Jacak}}\ and\ \bibinfo {author} {\bibfnamefont {Berndt}\
  \bibnamefont {Muller}},\ }\bibfield  {title} {\enquote {\bibinfo {title}
  {{The exploration of hot nuclear matter}},}\ }\href {\doibase
  10.1126/science.1215901} {\bibfield  {journal} {\bibinfo  {journal}
  {Science}\ }\textbf {\bibinfo {volume} {337}},\ \bibinfo {pages} {310--314}
  (\bibinfo {year} {2012})}\BibitemShut {NoStop}%
\bibitem [{\citenamefont {Gale}\ \emph {et~al.}(2022)\citenamefont {Gale},
  \citenamefont {Paquet}, \citenamefont {Schenke},\ and\ \citenamefont
  {Shen}}]{Gale:2021emg}%
  \BibitemOpen
  \bibfield  {author} {\bibinfo {author} {\bibfnamefont {Charles}\ \bibnamefont
  {Gale}}, \bibinfo {author} {\bibfnamefont {Jean-Fran\c{c}ois}\ \bibnamefont
  {Paquet}}, \bibinfo {author} {\bibfnamefont {Bj\"orn}\ \bibnamefont
  {Schenke}}, \ and\ \bibinfo {author} {\bibfnamefont {Chun}\ \bibnamefont
  {Shen}},\ }\bibfield  {title} {\enquote {\bibinfo {title} {{Multimessenger
  heavy-ion collision physics}},}\ }\href {\doibase
  10.1103/PhysRevC.105.014909} {\bibfield  {journal} {\bibinfo  {journal}
  {Phys. Rev. C}\ }\textbf {\bibinfo {volume} {105}},\ \bibinfo {pages}
  {014909} (\bibinfo {year} {2022})},\ \Eprint
  {http://arxiv.org/abs/2106.11216} {arXiv:2106.11216 [nucl-th]} \BibitemShut
  {NoStop}%
\bibitem [{\citenamefont {Paquet}\ \emph {et~al.}(2016)\citenamefont {Paquet},
  \citenamefont {Shen}, \citenamefont {Denicol}, \citenamefont {Luzum},
  \citenamefont {Schenke}, \citenamefont {Jeon},\ and\ \citenamefont
  {Gale}}]{Paquet:2015lta}%
  \BibitemOpen
  \bibfield  {author} {\bibinfo {author} {\bibfnamefont {Jean-Fran\c{c}ois}\
  \bibnamefont {Paquet}}, \bibinfo {author} {\bibfnamefont {Chun}\ \bibnamefont
  {Shen}}, \bibinfo {author} {\bibfnamefont {Gabriel~S.}\ \bibnamefont
  {Denicol}}, \bibinfo {author} {\bibfnamefont {Matthew}\ \bibnamefont
  {Luzum}}, \bibinfo {author} {\bibfnamefont {Bj{\"o}rn}\ \bibnamefont
  {Schenke}}, \bibinfo {author} {\bibfnamefont {Sangyong}\ \bibnamefont
  {Jeon}}, \ and\ \bibinfo {author} {\bibfnamefont {Charles}\ \bibnamefont
  {Gale}},\ }\bibfield  {title} {\enquote {\bibinfo {title} {{Production of
  photons in relativistic heavy-ion collisions}},}\ }\href {\doibase
  10.1103/PhysRevC.93.044906} {\bibfield  {journal} {\bibinfo  {journal} {Phys.
  Rev.}\ }\textbf {\bibinfo {volume} {C93}},\ \bibinfo {pages} {044906}
  (\bibinfo {year} {2016})},\ \Eprint {http://arxiv.org/abs/1509.06738}
  {arXiv:1509.06738 [hep-ph]} \BibitemShut {NoStop}%
\bibitem [{\citenamefont {van Hees}\ \emph {et~al.}(2011)\citenamefont {van
  Hees}, \citenamefont {Gale},\ and\ \citenamefont {Rapp}}]{vanHees:2011vb}%
  \BibitemOpen
  \bibfield  {author} {\bibinfo {author} {\bibfnamefont {Hendrik}\ \bibnamefont
  {van Hees}}, \bibinfo {author} {\bibfnamefont {Charles}\ \bibnamefont
  {Gale}}, \ and\ \bibinfo {author} {\bibfnamefont {Ralf}\ \bibnamefont
  {Rapp}},\ }\bibfield  {title} {\enquote {\bibinfo {title} {{Thermal Photons
  and Collective Flow at the Relativistic Heavy-Ion Collider}},}\ }\href
  {\doibase 10.1103/PhysRevC.84.054906} {\bibfield  {journal} {\bibinfo
  {journal} {Phys. Rev. C}\ }\textbf {\bibinfo {volume} {84}},\ \bibinfo
  {pages} {054906} (\bibinfo {year} {2011})},\ \Eprint
  {http://arxiv.org/abs/1108.2131} {arXiv:1108.2131 [hep-ph]} \BibitemShut
  {NoStop}%
\bibitem [{\citenamefont {Gale}\ and\ \citenamefont
  {Haglin}(2003)}]{Gale:2003iz}%
  \BibitemOpen
  \bibfield  {author} {\bibinfo {author} {\bibfnamefont {Charles}\ \bibnamefont
  {Gale}}\ and\ \bibinfo {author} {\bibfnamefont {Kevin~L.}\ \bibnamefont
  {Haglin}},\ }\bibfield  {title} {\enquote {\bibinfo {title} {{Electromagnetic
  radiation from relativistic nuclear collisions}},}\ }\href@noop {} {\ ,\
  \bibinfo {pages} {364--429} (\bibinfo {year} {2003})},\ \Eprint
  {http://arxiv.org/abs/hep-ph/0306098} {arXiv:hep-ph/0306098} \BibitemShut
  {NoStop}%
\bibitem [{\citenamefont {Gale}\ and\ \citenamefont
  {Kapusta}(1987)}]{Gale:1987ki}%
  \BibitemOpen
  \bibfield  {author} {\bibinfo {author} {\bibfnamefont {Charles}\ \bibnamefont
  {Gale}}\ and\ \bibinfo {author} {\bibfnamefont {Joseph~I.}\ \bibnamefont
  {Kapusta}},\ }\bibfield  {title} {\enquote {\bibinfo {title} {{Dilepton
  radiation from high temperature nuclear matter}},}\ }\href {\doibase
  10.1103/PhysRevC.35.2107} {\bibfield  {journal} {\bibinfo  {journal} {Phys.
  Rev. C}\ }\textbf {\bibinfo {volume} {35}},\ \bibinfo {pages} {2107--2116}
  (\bibinfo {year} {1987})}\BibitemShut {NoStop}%
\bibitem [{\citenamefont {van Hees}\ and\ \citenamefont
  {Rapp}(2008{\natexlab{a}})}]{vanHees:2007th}%
  \BibitemOpen
  \bibfield  {author} {\bibinfo {author} {\bibfnamefont {Hendrik}\ \bibnamefont
  {van Hees}}\ and\ \bibinfo {author} {\bibfnamefont {Ralf}\ \bibnamefont
  {Rapp}},\ }\bibfield  {title} {\enquote {\bibinfo {title} {{Dilepton
  Radiation at the CERN Super Proton Synchrotron}},}\ }\href {\doibase
  10.1016/j.nuclphysa.2008.03.009} {\bibfield  {journal} {\bibinfo  {journal}
  {Nucl. Phys. A}\ }\textbf {\bibinfo {volume} {806}},\ \bibinfo {pages}
  {339--387} (\bibinfo {year} {2008}{\natexlab{a}})},\ \Eprint
  {http://arxiv.org/abs/0711.3444} {arXiv:0711.3444 [hep-ph]} \BibitemShut
  {NoStop}%
\bibitem [{\citenamefont {Rapp}(2024)}]{Rapp:2024grb}%
  \BibitemOpen
  \bibfield  {author} {\bibinfo {author} {\bibfnamefont {Ralf}\ \bibnamefont
  {Rapp}},\ }\bibfield  {title} {\enquote {\bibinfo {title} {{Electric
  Conductivity of QCD Matter and Dilepton Spectra in Heavy-Ion Collisions}},}\
  }\href@noop {} {\  (\bibinfo {year} {2024})},\ \Eprint
  {http://arxiv.org/abs/2406.14656} {arXiv:2406.14656 [hep-ph]} \BibitemShut
  {NoStop}%
\bibitem [{\citenamefont {Wu}\ \emph {et~al.}(2024)\citenamefont {Wu},
  \citenamefont {Du}, \citenamefont {Gale},\ and\ \citenamefont
  {Jeon}}]{Wu:2024pba}%
  \BibitemOpen
  \bibfield  {author} {\bibinfo {author} {\bibfnamefont {Xiang-Yu}\
  \bibnamefont {Wu}}, \bibinfo {author} {\bibfnamefont {Lipei}\ \bibnamefont
  {Du}}, \bibinfo {author} {\bibfnamefont {Charles}\ \bibnamefont {Gale}}, \
  and\ \bibinfo {author} {\bibfnamefont {Sangyong}\ \bibnamefont {Jeon}},\
  }\bibfield  {title} {\enquote {\bibinfo {title} {{Probing the equilibration
  of the QCD matter created in heavy-ion collisions with dileptons}},}\ }\href
  {\doibase 10.1103/PhysRevC.110.054904} {\bibfield  {journal} {\bibinfo
  {journal} {Phys. Rev. C}\ }\textbf {\bibinfo {volume} {110}},\ \bibinfo
  {pages} {054904} (\bibinfo {year} {2024})},\ \Eprint
  {http://arxiv.org/abs/2407.04156} {arXiv:2407.04156 [nucl-th]} \BibitemShut
  {NoStop}%
\bibitem [{\citenamefont {Garcia-Montero}\ \emph {et~al.}(2024)\citenamefont
  {Garcia-Montero}, \citenamefont {Plaschke},\ and\ \citenamefont
  {Schlichting}}]{Garcia-Montero:2024lbl}%
  \BibitemOpen
  \bibfield  {author} {\bibinfo {author} {\bibfnamefont {Oscar}\ \bibnamefont
  {Garcia-Montero}}, \bibinfo {author} {\bibfnamefont {Philip}\ \bibnamefont
  {Plaschke}}, \ and\ \bibinfo {author} {\bibfnamefont {S\"oren}\ \bibnamefont
  {Schlichting}},\ }\bibfield  {title} {\enquote {\bibinfo {title} {{Scaling of
  pre-equilibrium dilepton production in QCD Kinetic Theory}},}\ }\href@noop {}
  {\  (\bibinfo {year} {2024})},\ \Eprint {http://arxiv.org/abs/2403.04846}
  {arXiv:2403.04846 [hep-ph]} \BibitemShut {NoStop}%
\bibitem [{\citenamefont {Coquet}\ \emph {et~al.}(2021)\citenamefont {Coquet},
  \citenamefont {Du}, \citenamefont {Ollitrault}, \citenamefont {Schlichting},\
  and\ \citenamefont {Winn}}]{Coquet:2021lca}%
  \BibitemOpen
  \bibfield  {author} {\bibinfo {author} {\bibfnamefont {Maurice}\ \bibnamefont
  {Coquet}}, \bibinfo {author} {\bibfnamefont {Xiaojian}\ \bibnamefont {Du}},
  \bibinfo {author} {\bibfnamefont {Jean-Yves}\ \bibnamefont {Ollitrault}},
  \bibinfo {author} {\bibfnamefont {Soeren}\ \bibnamefont {Schlichting}}, \
  and\ \bibinfo {author} {\bibfnamefont {Michael}\ \bibnamefont {Winn}},\
  }\bibfield  {title} {\enquote {\bibinfo {title} {{Intermediate mass dileptons
  as pre-equilibrium probes in heavy ion collisions}},}\ }\href {\doibase
  10.1016/j.physletb.2021.136626} {\bibfield  {journal} {\bibinfo  {journal}
  {Phys. Lett. B}\ }\textbf {\bibinfo {volume} {821}},\ \bibinfo {pages}
  {136626} (\bibinfo {year} {2021})},\ \Eprint
  {http://arxiv.org/abs/2104.07622} {arXiv:2104.07622 [nucl-th]} \BibitemShut
  {NoStop}%
\bibitem [{\citenamefont {Wang}\ \emph {et~al.}(2020)\citenamefont {Wang},
  \citenamefont {Shovkovy}, \citenamefont {Yu},\ and\ \citenamefont
  {Huang}}]{Wang:2020dsr}%
  \BibitemOpen
  \bibfield  {author} {\bibinfo {author} {\bibfnamefont {Xinyang}\ \bibnamefont
  {Wang}}, \bibinfo {author} {\bibfnamefont {Igor~A.}\ \bibnamefont
  {Shovkovy}}, \bibinfo {author} {\bibfnamefont {Lang}\ \bibnamefont {Yu}}, \
  and\ \bibinfo {author} {\bibfnamefont {Mei}\ \bibnamefont {Huang}},\
  }\bibfield  {title} {\enquote {\bibinfo {title} {{Ellipticity of photon
  emission from strongly magnetized hot QCD plasma}},}\ }\href {\doibase
  10.1103/PhysRevD.102.076010} {\bibfield  {journal} {\bibinfo  {journal}
  {Phys. Rev. D}\ }\textbf {\bibinfo {volume} {102}},\ \bibinfo {pages}
  {076010} (\bibinfo {year} {2020})},\ \Eprint
  {http://arxiv.org/abs/2006.16254} {arXiv:2006.16254 [hep-ph]} \BibitemShut
  {NoStop}%
\bibitem [{\citenamefont {Wang}\ and\ \citenamefont
  {Shovkovy}(2021)}]{Wang:2021ebh}%
  \BibitemOpen
  \bibfield  {author} {\bibinfo {author} {\bibfnamefont {Xinyang}\ \bibnamefont
  {Wang}}\ and\ \bibinfo {author} {\bibfnamefont {Igor}\ \bibnamefont
  {Shovkovy}},\ }\bibfield  {title} {\enquote {\bibinfo {title} {{Photon
  polarization tensor in a magnetized plasma: Absorptive part}},}\ }\href
  {\doibase 10.1103/PhysRevD.104.056017} {\bibfield  {journal} {\bibinfo
  {journal} {Phys. Rev. D}\ }\textbf {\bibinfo {volume} {104}},\ \bibinfo
  {pages} {056017} (\bibinfo {year} {2021})},\ \Eprint
  {http://arxiv.org/abs/2103.01967} {arXiv:2103.01967 [nucl-th]} \BibitemShut
  {NoStop}%
\bibitem [{\citenamefont {Wang}\ and\ \citenamefont
  {Shovkovy}(2022)}]{Wang:2022jxx}%
  \BibitemOpen
  \bibfield  {author} {\bibinfo {author} {\bibfnamefont {Xinyang}\ \bibnamefont
  {Wang}}\ and\ \bibinfo {author} {\bibfnamefont {Igor~A.}\ \bibnamefont
  {Shovkovy}},\ }\bibfield  {title} {\enquote {\bibinfo {title} {{Rate and
  ellipticity of dilepton production in a magnetized quark-gluon plasma}},}\
  }\href {\doibase 10.1103/PhysRevD.106.036014} {\bibfield  {journal} {\bibinfo
   {journal} {Phys. Rev. D}\ }\textbf {\bibinfo {volume} {106}},\ \bibinfo
  {pages} {036014} (\bibinfo {year} {2022})},\ \Eprint
  {http://arxiv.org/abs/2205.00276} {arXiv:2205.00276 [nucl-th]} \BibitemShut
  {NoStop}%
\bibitem [{\citenamefont {Kimura}\ \emph {et~al.}(2024)\citenamefont {Kimura},
  \citenamefont {Benoit}, \citenamefont {Ishikawa}, \citenamefont {Nonaka},\
  and\ \citenamefont {Shigaki}}]{Kimura:2024gao}%
  \BibitemOpen
  \bibfield  {author} {\bibinfo {author} {\bibfnamefont {Kento}\ \bibnamefont
  {Kimura}}, \bibinfo {author} {\bibfnamefont {Nicholas~J.}\ \bibnamefont
  {Benoit}}, \bibinfo {author} {\bibfnamefont {Ken-Ichi}\ \bibnamefont
  {Ishikawa}}, \bibinfo {author} {\bibfnamefont {Chiho}\ \bibnamefont
  {Nonaka}}, \ and\ \bibinfo {author} {\bibfnamefont {Kenta}\ \bibnamefont
  {Shigaki}},\ }\bibfield  {title} {\enquote {\bibinfo {title} {{Estimate of
  virtual photon polarization due to the intense magnetic field in Pb-Pb
  collisions at the LHC energies}},}\ }\href@noop {} {\  (\bibinfo {year}
  {2024})},\ \Eprint {http://arxiv.org/abs/2411.01406} {arXiv:2411.01406
  [hep-ph]} \BibitemShut {NoStop}%
\bibitem [{\citenamefont {Arnaldi}\ \emph
  {et~al.}(2009{\natexlab{a}})\citenamefont {Arnaldi} \emph
  {et~al.}}]{NA60:2008ctj}%
  \BibitemOpen
  \bibfield  {author} {\bibinfo {author} {\bibfnamefont {R.}~\bibnamefont
  {Arnaldi}} \emph {et~al.} (\bibinfo {collaboration} {NA60}),\ }\bibfield
  {title} {\enquote {\bibinfo {title} {{NA60 results on thermal dimuons}},}\
  }\href {\doibase 10.1140/epjc/s10052-009-0878-5} {\bibfield  {journal}
  {\bibinfo  {journal} {Eur. Phys. J. C}\ }\textbf {\bibinfo {volume} {61}},\
  \bibinfo {pages} {711--720} (\bibinfo {year} {2009}{\natexlab{a}})},\ \Eprint
  {http://arxiv.org/abs/0812.3053} {arXiv:0812.3053 [nucl-ex]} \BibitemShut
  {NoStop}%
\bibitem [{\citenamefont {Churchill}\ \emph
  {et~al.}(2024{\natexlab{a}})\citenamefont {Churchill}, \citenamefont {Du},
  \citenamefont {Gale}, \citenamefont {Jackson},\ and\ \citenamefont
  {Jeon}}]{Churchill:2023zkk}%
  \BibitemOpen
  \bibfield  {author} {\bibinfo {author} {\bibfnamefont {Jessica}\ \bibnamefont
  {Churchill}}, \bibinfo {author} {\bibfnamefont {Lipei}\ \bibnamefont {Du}},
  \bibinfo {author} {\bibfnamefont {Charles}\ \bibnamefont {Gale}}, \bibinfo
  {author} {\bibfnamefont {Greg}\ \bibnamefont {Jackson}}, \ and\ \bibinfo
  {author} {\bibfnamefont {Sangyong}\ \bibnamefont {Jeon}},\ }\bibfield
  {title} {\enquote {\bibinfo {title} {{Virtual Photons Shed Light on the Early
  Temperature of Dense QCD Matter}},}\ }\href {\doibase
  10.1103/PhysRevLett.132.172301} {\bibfield  {journal} {\bibinfo  {journal}
  {Phys. Rev. Lett.}\ }\textbf {\bibinfo {volume} {132}},\ \bibinfo {pages}
  {172301} (\bibinfo {year} {2024}{\natexlab{a}})},\ \Eprint
  {http://arxiv.org/abs/2311.06951} {arXiv:2311.06951 [nucl-th]} \BibitemShut
  {NoStop}%
\bibitem [{\citenamefont {Churchill}\ \emph
  {et~al.}(2024{\natexlab{b}})\citenamefont {Churchill}, \citenamefont {Du},
  \citenamefont {Gale}, \citenamefont {Jackson},\ and\ \citenamefont
  {Jeon}}]{Churchill:2023vpt}%
  \BibitemOpen
  \bibfield  {author} {\bibinfo {author} {\bibfnamefont {Jessica}\ \bibnamefont
  {Churchill}}, \bibinfo {author} {\bibfnamefont {Lipei}\ \bibnamefont {Du}},
  \bibinfo {author} {\bibfnamefont {Charles}\ \bibnamefont {Gale}}, \bibinfo
  {author} {\bibfnamefont {Greg}\ \bibnamefont {Jackson}}, \ and\ \bibinfo
  {author} {\bibfnamefont {Sangyong}\ \bibnamefont {Jeon}},\ }\bibfield
  {title} {\enquote {\bibinfo {title} {{Dilepton production at next-to-leading
  order and intermediate invariant-mass observables}},}\ }\href {\doibase
  10.1103/PhysRevC.109.044915} {\bibfield  {journal} {\bibinfo  {journal}
  {Phys. Rev. C}\ }\textbf {\bibinfo {volume} {109}},\ \bibinfo {pages}
  {044915} (\bibinfo {year} {2024}{\natexlab{b}})},\ \Eprint
  {http://arxiv.org/abs/2311.06675} {arXiv:2311.06675 [nucl-th]} \BibitemShut
  {NoStop}%
\bibitem [{\citenamefont {Acharya}\ \emph {et~al.}(2023)\citenamefont {Acharya}
  \emph {et~al.}}]{ALICE:2023jef}%
  \BibitemOpen
  \bibfield  {author} {\bibinfo {author} {\bibfnamefont {Shreyasi}\
  \bibnamefont {Acharya}} \emph {et~al.} (\bibinfo {collaboration} {ALICE}),\
  }\bibfield  {title} {\enquote {\bibinfo {title} {{Dielectron production in
  central Pb$-$Pb collisions at $\sqrt{s_\mathrm{NN}}$ = 5.02 TeV}},}\
  }\href@noop {} {\  (\bibinfo {year} {2023})},\ \Eprint
  {http://arxiv.org/abs/2308.16704} {arXiv:2308.16704 [nucl-ex]} \BibitemShut
  {NoStop}%
\bibitem [{\citenamefont {Abdulhamid}\ \emph {et~al.}(2023)\citenamefont
  {Abdulhamid} \emph {et~al.}}]{STAR:2023wta}%
  \BibitemOpen
  \bibfield  {author} {\bibinfo {author} {\bibfnamefont {M.~I.}\ \bibnamefont
  {Abdulhamid}} \emph {et~al.} (\bibinfo {collaboration} {STAR}),\ }\bibfield
  {title} {\enquote {\bibinfo {title} {{Measurements of dielectron production
  in Au+Au collisions at $\sqrt{s_{\rm NN}}=27$, 39, and 62.4 GeV from the STAR
  experiment}},}\ }\href {\doibase 10.1103/PhysRevC.107.L061901} {\bibfield
  {journal} {\bibinfo  {journal} {Phys. Rev. C}\ }\textbf {\bibinfo {volume}
  {107}},\ \bibinfo {pages} {L061901} (\bibinfo {year} {2023})}\BibitemShut
  {NoStop}%
\bibitem [{\citenamefont {Adamczyk}\ \emph {et~al.}(2015)\citenamefont
  {Adamczyk} \emph {et~al.}}]{STAR:2015tnn}%
  \BibitemOpen
  \bibfield  {author} {\bibinfo {author} {\bibfnamefont {L.}~\bibnamefont
  {Adamczyk}} \emph {et~al.} (\bibinfo {collaboration} {STAR}),\ }\bibfield
  {title} {\enquote {\bibinfo {title} {{Measurements of Dielectron Production
  in Au$+$Au Collisions at $\sqrt{s_{\rm NN}}$ = 200 GeV from the STAR
  Experiment}},}\ }\href {\doibase 10.1103/PhysRevC.92.024912} {\bibfield
  {journal} {\bibinfo  {journal} {Phys. Rev. C}\ }\textbf {\bibinfo {volume}
  {92}},\ \bibinfo {pages} {024912} (\bibinfo {year} {2015})},\ \Eprint
  {http://arxiv.org/abs/1504.01317} {arXiv:1504.01317 [hep-ex]} \BibitemShut
  {NoStop}%
\bibitem [{STA(2024)}]{STAR:2024bpc}%
  \BibitemOpen
  \bibfield  {title} {\enquote {\bibinfo {title} {{Temperature Measurement of
  Quark-Gluon Plasma at Different Stages}},}\ }\href@noop {} {\  (\bibinfo
  {year} {2024})},\ \Eprint {http://arxiv.org/abs/2402.01998} {arXiv:2402.01998
  [nucl-ex]} \BibitemShut {NoStop}%
\bibitem [{\citenamefont {Arnaldi}\ \emph
  {et~al.}(2009{\natexlab{b}})\citenamefont {Arnaldi} \emph
  {et~al.}}]{NA60:2008dcb}%
  \BibitemOpen
  \bibfield  {author} {\bibinfo {author} {\bibfnamefont {R}~\bibnamefont
  {Arnaldi}} \emph {et~al.} (\bibinfo {collaboration} {NA60}),\ }\bibfield
  {title} {\enquote {\bibinfo {title} {{Evidence for the production of
  thermal-like muon pairs with masses above 1~GeV/$c^2$ in 158~$A$~GeV
  Indium-Indium Collisions}},}\ }\href {\doibase
  10.1140/epjc/s10052-008-0857-2} {\bibfield  {journal} {\bibinfo  {journal}
  {Eur. Phys. J. C}\ }\textbf {\bibinfo {volume} {59}},\ \bibinfo {pages}
  {607--623} (\bibinfo {year} {2009}{\natexlab{b}})},\ \Eprint
  {http://arxiv.org/abs/0810.3204} {arXiv:0810.3204 [nucl-ex]} \BibitemShut
  {NoStop}%
\bibitem [{\citenamefont {Rapp}(2013)}]{Rapp:2013nxa}%
  \BibitemOpen
  \bibfield  {author} {\bibinfo {author} {\bibfnamefont {Ralf}\ \bibnamefont
  {Rapp}},\ }\bibfield  {title} {\enquote {\bibinfo {title} {{Dilepton
  Spectroscopy of QCD Matter at Collider Energies}},}\ }\href {\doibase
  10.1155/2013/148253} {\bibfield  {journal} {\bibinfo  {journal} {Adv. High
  Energy Phys.}\ }\textbf {\bibinfo {volume} {2013}},\ \bibinfo {pages}
  {148253} (\bibinfo {year} {2013})},\ \Eprint {http://arxiv.org/abs/1304.2309}
  {arXiv:1304.2309 [hep-ph]} \BibitemShut {NoStop}%
\bibitem [{Note1()}]{Note1}%
  \BibitemOpen
  \bibinfo {note} {This channel is also included in the Drell-Yan process where
  initial quarks are sourced from the nPDFs of the colliding ions.}\BibitemShut
  {Stop}%
\bibitem [{\citenamefont {Laine}(2013)}]{Laine:2013vma}%
  \BibitemOpen
  \bibfield  {author} {\bibinfo {author} {\bibfnamefont {M.}~\bibnamefont
  {Laine}},\ }\bibfield  {title} {\enquote {\bibinfo {title} {{NLO thermal
  dilepton rate at non-zero momentum}},}\ }\href {\doibase
  10.1007/JHEP11(2013)120} {\bibfield  {journal} {\bibinfo  {journal} {JHEP}\
  }\textbf {\bibinfo {volume} {11}},\ \bibinfo {pages} {120} (\bibinfo {year}
  {2013})},\ \Eprint {http://arxiv.org/abs/1310.0164} {arXiv:1310.0164
  [hep-ph]} \BibitemShut {NoStop}%
\bibitem [{\citenamefont {Jackson}(2019)}]{Jackson:2019mop}%
  \BibitemOpen
  \bibfield  {author} {\bibinfo {author} {\bibfnamefont {G.}~\bibnamefont
  {Jackson}},\ }\bibfield  {title} {\enquote {\bibinfo {title} {{Two-loop
  thermal spectral functions with general kinematics}},}\ }\href {\doibase
  10.1103/PhysRevD.100.116019} {\bibfield  {journal} {\bibinfo  {journal}
  {Phys. Rev. D}\ }\textbf {\bibinfo {volume} {100}},\ \bibinfo {pages}
  {116019} (\bibinfo {year} {2019})},\ \Eprint
  {http://arxiv.org/abs/1910.07552} {arXiv:1910.07552 [hep-ph]} \BibitemShut
  {NoStop}%
\bibitem [{\citenamefont {Arnold}\ \emph
  {et~al.}(2001{\natexlab{a}})\citenamefont {Arnold}, \citenamefont {Moore},\
  and\ \citenamefont {Yaffe}}]{Arnold:2001ba}%
  \BibitemOpen
  \bibfield  {author} {\bibinfo {author} {\bibfnamefont {Peter~Brockway}\
  \bibnamefont {Arnold}}, \bibinfo {author} {\bibfnamefont {Guy~D.}\
  \bibnamefont {Moore}}, \ and\ \bibinfo {author} {\bibfnamefont {Laurence~G.}\
  \bibnamefont {Yaffe}},\ }\bibfield  {title} {\enquote {\bibinfo {title}
  {{Photon emission from ultrarelativistic plasmas}},}\ }\href {\doibase
  10.1088/1126-6708/2001/11/057} {\bibfield  {journal} {\bibinfo  {journal}
  {JHEP}\ }\textbf {\bibinfo {volume} {11}},\ \bibinfo {pages} {057} (\bibinfo
  {year} {2001}{\natexlab{a}})},\ \Eprint {http://arxiv.org/abs/hep-ph/0109064}
  {arXiv:hep-ph/0109064} \BibitemShut {NoStop}%
\bibitem [{\citenamefont {Arnold}\ \emph
  {et~al.}(2001{\natexlab{b}})\citenamefont {Arnold}, \citenamefont {Moore},\
  and\ \citenamefont {Yaffe}}]{Arnold:2001ms}%
  \BibitemOpen
  \bibfield  {author} {\bibinfo {author} {\bibfnamefont {Peter~Brockway}\
  \bibnamefont {Arnold}}, \bibinfo {author} {\bibfnamefont {Guy~D.}\
  \bibnamefont {Moore}}, \ and\ \bibinfo {author} {\bibfnamefont {Laurence~G.}\
  \bibnamefont {Yaffe}},\ }\bibfield  {title} {\enquote {\bibinfo {title}
  {{Photon emission from quark gluon plasma: Complete leading order
  results}},}\ }\href {\doibase 10.1088/1126-6708/2001/12/009} {\bibfield
  {journal} {\bibinfo  {journal} {JHEP}\ }\textbf {\bibinfo {volume} {12}},\
  \bibinfo {pages} {009} (\bibinfo {year} {2001}{\natexlab{b}})},\ \Eprint
  {http://arxiv.org/abs/hep-ph/0111107} {arXiv:hep-ph/0111107 [hep-ph]}
  \BibitemShut {NoStop}%
\bibitem [{\citenamefont {Aurenche}\ \emph
  {et~al.}(2002{\natexlab{a}})\citenamefont {Aurenche}, \citenamefont {Gelis},\
  and\ \citenamefont {Zaraket}}]{Aurenche:2002pc}%
  \BibitemOpen
  \bibfield  {author} {\bibinfo {author} {\bibfnamefont {P.}~\bibnamefont
  {Aurenche}}, \bibinfo {author} {\bibfnamefont {F.}~\bibnamefont {Gelis}}, \
  and\ \bibinfo {author} {\bibfnamefont {H.}~\bibnamefont {Zaraket}},\
  }\bibfield  {title} {\enquote {\bibinfo {title} {{Enhanced thermal production
  of hard dileptons by $3\to 2$ processes}},}\ }\href {\doibase
  10.1088/1126-6708/2002/07/063} {\bibfield  {journal} {\bibinfo  {journal}
  {JHEP}\ }\textbf {\bibinfo {volume} {07}},\ \bibinfo {pages} {063} (\bibinfo
  {year} {2002}{\natexlab{a}})},\ \Eprint {http://arxiv.org/abs/hep-ph/0204145}
  {arXiv:hep-ph/0204145} \BibitemShut {NoStop}%
\bibitem [{\citenamefont {Aurenche}\ \emph
  {et~al.}(2002{\natexlab{b}})\citenamefont {Aurenche}, \citenamefont {Gelis},
  \citenamefont {Moore},\ and\ \citenamefont {Zaraket}}]{Aurenche:2002wq}%
  \BibitemOpen
  \bibfield  {author} {\bibinfo {author} {\bibfnamefont {P.}~\bibnamefont
  {Aurenche}}, \bibinfo {author} {\bibfnamefont {F.}~\bibnamefont {Gelis}},
  \bibinfo {author} {\bibfnamefont {G.~D.}\ \bibnamefont {Moore}}, \ and\
  \bibinfo {author} {\bibfnamefont {H.}~\bibnamefont {Zaraket}},\ }\bibfield
  {title} {\enquote {\bibinfo {title} {{Landau-Pomeranchuk-Migdal resummation
  for dilepton production}},}\ }\href {\doibase 10.1088/1126-6708/2002/12/006}
  {\bibfield  {journal} {\bibinfo  {journal} {JHEP}\ }\textbf {\bibinfo
  {volume} {12}},\ \bibinfo {pages} {006} (\bibinfo {year}
  {2002}{\natexlab{b}})},\ \Eprint {http://arxiv.org/abs/hep-ph/0211036}
  {arXiv:hep-ph/0211036 [hep-ph]} \BibitemShut {NoStop}%
\bibitem [{\citenamefont {Ghisoiu}\ and\ \citenamefont
  {Laine}(2014)}]{Ghisoiu:2014mha}%
  \BibitemOpen
  \bibfield  {author} {\bibinfo {author} {\bibfnamefont {I.}~\bibnamefont
  {Ghisoiu}}\ and\ \bibinfo {author} {\bibfnamefont {M.}~\bibnamefont
  {Laine}},\ }\bibfield  {title} {\enquote {\bibinfo {title} {{Interpolation of
  hard and soft dilepton rates}},}\ }\href {\doibase 10.1007/JHEP10(2014)083}
  {\bibfield  {journal} {\bibinfo  {journal} {JHEP}\ }\textbf {\bibinfo
  {volume} {10}},\ \bibinfo {pages} {083} (\bibinfo {year} {2014})},\ \Eprint
  {http://arxiv.org/abs/1407.7955} {arXiv:1407.7955 [hep-ph]} \BibitemShut
  {NoStop}%
\bibitem [{\citenamefont {Jackson}\ and\ \citenamefont
  {Laine}(2019)}]{Jackson:2019yao}%
  \BibitemOpen
  \bibfield  {author} {\bibinfo {author} {\bibfnamefont {G.}~\bibnamefont
  {Jackson}}\ and\ \bibinfo {author} {\bibfnamefont {M.}~\bibnamefont
  {Laine}},\ }\bibfield  {title} {\enquote {\bibinfo {title} {{Testing thermal
  photon and dilepton rates}},}\ }\href {\doibase 10.1007/JHEP11(2019)144}
  {\bibfield  {journal} {\bibinfo  {journal} {JHEP}\ }\textbf {\bibinfo
  {volume} {11}},\ \bibinfo {pages} {144} (\bibinfo {year} {2019})},\ \Eprint
  {http://arxiv.org/abs/1910.09567} {arXiv:1910.09567 [hep-ph]} \BibitemShut
  {NoStop}%
\bibitem [{\citenamefont {Ghiglieri}\ and\ \citenamefont
  {Laine}(2022)}]{Ghiglieri:2021vcq}%
  \BibitemOpen
  \bibfield  {author} {\bibinfo {author} {\bibfnamefont {J.}~\bibnamefont
  {Ghiglieri}}\ and\ \bibinfo {author} {\bibfnamefont {M.}~\bibnamefont
  {Laine}},\ }\bibfield  {title} {\enquote {\bibinfo {title} {{Smooth
  interpolation between thermal Born and LPM rates}},}\ }\href {\doibase
  10.1007/JHEP01(2022)173} {\bibfield  {journal} {\bibinfo  {journal} {JHEP}\
  }\textbf {\bibinfo {volume} {01}},\ \bibinfo {pages} {173} (\bibinfo {year}
  {2022})},\ \Eprint {http://arxiv.org/abs/2110.07149} {arXiv:2110.07149
  [hep-ph]} \BibitemShut {NoStop}%
\bibitem [{Note2()}]{Note2}%
  \BibitemOpen
  \bibinfo {note} {From this point on, we consider the ``NLO contributions'' to
  include the LPM resummation for small $M$ as in Refs.~\cite
  {Ghisoiu:2014mha,Jackson:2019yao}.}\BibitemShut {Stop}%
\bibitem [{\citenamefont {Churchill}\ \emph
  {et~al.}(2024{\natexlab{c}})\citenamefont {Churchill}, \citenamefont {Du},
  \citenamefont {Forster}, \citenamefont {Gao}, \citenamefont {Jackson},
  \citenamefont {Jeon},\ and\ \citenamefont {Gale}}]{Churchill:2023hog}%
  \BibitemOpen
  \bibfield  {author} {\bibinfo {author} {\bibfnamefont {Jessica}\ \bibnamefont
  {Churchill}}, \bibinfo {author} {\bibfnamefont {Lipei}\ \bibnamefont {Du}},
  \bibinfo {author} {\bibfnamefont {Bailey}\ \bibnamefont {Forster}}, \bibinfo
  {author} {\bibfnamefont {Han}\ \bibnamefont {Gao}}, \bibinfo {author}
  {\bibfnamefont {Greg}\ \bibnamefont {Jackson}}, \bibinfo {author}
  {\bibfnamefont {Sangyong}\ \bibnamefont {Jeon}}, \ and\ \bibinfo {author}
  {\bibfnamefont {Charles}\ \bibnamefont {Gale}},\ }\bibfield  {title}
  {\enquote {\bibinfo {title} {{Thermal dilepton production in heavy-ion
  collisions at beam-energy-scan (BES) energies}},}\ }\href {\doibase
  10.1051/epjconf/202429607006} {\bibfield  {journal} {\bibinfo  {journal} {EPJ
  Web Conf.}\ }\textbf {\bibinfo {volume} {296}},\ \bibinfo {pages} {07006}
  (\bibinfo {year} {2024}{\natexlab{c}})},\ \Eprint
  {http://arxiv.org/abs/2312.10166} {arXiv:2312.10166 [nucl-th]} \BibitemShut
  {NoStop}%
\bibitem [{\citenamefont {Speranza}\ \emph {et~al.}(2018)\citenamefont
  {Speranza}, \citenamefont {Jaiswal},\ and\ \citenamefont
  {Friman}}]{Speranza:2018osi}%
  \BibitemOpen
  \bibfield  {author} {\bibinfo {author} {\bibfnamefont {Enrico}\ \bibnamefont
  {Speranza}}, \bibinfo {author} {\bibfnamefont {Amaresh}\ \bibnamefont
  {Jaiswal}}, \ and\ \bibinfo {author} {\bibfnamefont {Bengt}\ \bibnamefont
  {Friman}},\ }\bibfield  {title} {\enquote {\bibinfo {title} {{Virtual photon
  polarization and dilepton anisotropy in relativistic
  nucleus\textendash{}nucleus collisions}},}\ }\href {\doibase
  10.1016/j.physletb.2018.05.053} {\bibfield  {journal} {\bibinfo  {journal}
  {Phys. Lett. B}\ }\textbf {\bibinfo {volume} {782}},\ \bibinfo {pages}
  {395--400} (\bibinfo {year} {2018})},\ \Eprint
  {http://arxiv.org/abs/1802.02479} {arXiv:1802.02479 [hep-ph]} \BibitemShut
  {NoStop}%
\bibitem [{\citenamefont {Wei}\ and\ \citenamefont {Yan}(2024)}]{Wei:2024lah}%
  \BibitemOpen
  \bibfield  {author} {\bibinfo {author} {\bibfnamefont {Minghua}\ \bibnamefont
  {Wei}}\ and\ \bibinfo {author} {\bibfnamefont {Li}~\bibnamefont {Yan}},\
  }\bibfield  {title} {\enquote {\bibinfo {title} {{Weak magnetic field effect
  on dilepton polarization in heavy-ion collisions}},}\ }\href {\doibase
  10.1103/PhysRevD.110.054024} {\bibfield  {journal} {\bibinfo  {journal}
  {Phys. Rev. D}\ }\textbf {\bibinfo {volume} {110}},\ \bibinfo {pages}
  {054024} (\bibinfo {year} {2024})},\ \Eprint
  {http://arxiv.org/abs/2406.10041} {arXiv:2406.10041 [nucl-th]} \BibitemShut
  {NoStop}%
\bibitem [{\citenamefont {Seck}\ \emph {et~al.}(2023)\citenamefont {Seck},
  \citenamefont {Friman}, \citenamefont {Galatyuk}, \citenamefont {van Hees},
  \citenamefont {Speranza}, \citenamefont {Rapp},\ and\ \citenamefont
  {Wambach}}]{Seck:2023oyt}%
  \BibitemOpen
  \bibfield  {author} {\bibinfo {author} {\bibfnamefont {Florian}\ \bibnamefont
  {Seck}}, \bibinfo {author} {\bibfnamefont {Bengt}\ \bibnamefont {Friman}},
  \bibinfo {author} {\bibfnamefont {Tetyana}\ \bibnamefont {Galatyuk}},
  \bibinfo {author} {\bibfnamefont {Hendrik}\ \bibnamefont {van Hees}},
  \bibinfo {author} {\bibfnamefont {Enrico}\ \bibnamefont {Speranza}}, \bibinfo
  {author} {\bibfnamefont {Ralf}\ \bibnamefont {Rapp}}, \ and\ \bibinfo
  {author} {\bibfnamefont {Jochen}\ \bibnamefont {Wambach}},\ }\bibfield
  {title} {\enquote {\bibinfo {title} {{Polarization of Thermal Dilepton
  Radiation}},}\ }\href@noop {} {\  (\bibinfo {year} {2023})},\ \Eprint
  {http://arxiv.org/abs/2309.03189} {arXiv:2309.03189 [nucl-th]} \BibitemShut
  {NoStop}%
\bibitem [{Note3()}]{Note3}%
  \BibitemOpen
  \bibinfo {note} {Also the centre-of-mass frame of the dilepton.}\BibitemShut
  {Stop}%
\bibitem [{\citenamefont {Coquet}\ \emph {et~al.}(2024)\citenamefont {Coquet},
  \citenamefont {Winn}, \citenamefont {Du}, \citenamefont {Ollitrault},\ and\
  \citenamefont {Schlichting}}]{Coquet:2023wjk}%
  \BibitemOpen
  \bibfield  {author} {\bibinfo {author} {\bibfnamefont {Maurice}\ \bibnamefont
  {Coquet}}, \bibinfo {author} {\bibfnamefont {Michael}\ \bibnamefont {Winn}},
  \bibinfo {author} {\bibfnamefont {Xiaojian}\ \bibnamefont {Du}}, \bibinfo
  {author} {\bibfnamefont {Jean-Yves}\ \bibnamefont {Ollitrault}}, \ and\
  \bibinfo {author} {\bibfnamefont {Soeren}\ \bibnamefont {Schlichting}},\
  }\bibfield  {title} {\enquote {\bibinfo {title} {{Dilepton Polarization as a
  Signature of Plasma Anisotropy}},}\ }\href {\doibase
  10.1103/PhysRevLett.132.232301} {\bibfield  {journal} {\bibinfo  {journal}
  {Phys. Rev. Lett.}\ }\textbf {\bibinfo {volume} {132}},\ \bibinfo {pages}
  {232301} (\bibinfo {year} {2024})},\ \Eprint
  {http://arxiv.org/abs/2309.00555} {arXiv:2309.00555 [nucl-th]} \BibitemShut
  {NoStop}%
\bibitem [{\citenamefont {Hauksson}\ and\ \citenamefont
  {Gale}(2024)}]{Hauksson:2023dwh}%
  \BibitemOpen
  \bibfield  {author} {\bibinfo {author} {\bibfnamefont {Sigtryggur}\
  \bibnamefont {Hauksson}}\ and\ \bibinfo {author} {\bibfnamefont {Charles}\
  \bibnamefont {Gale}},\ }\bibfield  {title} {\enquote {\bibinfo {title}
  {{Polarized photons from the early stages of relativistic heavy-ion
  collisions}},}\ }\href {\doibase 10.1103/PhysRevC.109.034902} {\bibfield
  {journal} {\bibinfo  {journal} {Phys. Rev. C}\ }\textbf {\bibinfo {volume}
  {109}},\ \bibinfo {pages} {034902} (\bibinfo {year} {2024})},\ \Eprint
  {http://arxiv.org/abs/2306.10307} {arXiv:2306.10307 [nucl-th]} \BibitemShut
  {NoStop}%
\bibitem [{Note4()}]{Note4}%
  \BibitemOpen
  \bibinfo {note} {The temperature is defined in the fluid rest
  frame.}\BibitemShut {Stop}%
\bibitem [{\citenamefont {Kapusta}\ and\ \citenamefont
  {Gale}(2023)}]{Kapusta:2023eix}%
  \BibitemOpen
  \bibfield  {author} {\bibinfo {author} {\bibfnamefont {Joseph~I.}\
  \bibnamefont {Kapusta}}\ and\ \bibinfo {author} {\bibfnamefont {Charles}\
  \bibnamefont {Gale}},\ }\href {\doibase 10.1017/9781009401968} {\emph
  {\bibinfo {title} {{Finite-Temperature Field Theory}}}},\ Cambridge
  Monographs on Mathematical Physics\ (\bibinfo  {publisher} {Cambridge
  University Press},\ \bibinfo {year} {2023})\BibitemShut {NoStop}%
\bibitem [{\citenamefont {Laine}\ and\ \citenamefont
  {Vuorinen}(2016)}]{Laine:2016hma}%
  \BibitemOpen
  \bibfield  {author} {\bibinfo {author} {\bibfnamefont {Mikko}\ \bibnamefont
  {Laine}}\ and\ \bibinfo {author} {\bibfnamefont {Aleksi}\ \bibnamefont
  {Vuorinen}},\ }\href {\doibase 10.1007/978-3-319-31933-9} {\emph {\bibinfo
  {title} {{Basics of Thermal Field Theory}}}},\ Vol.\ \bibinfo {volume} {925}\
  (\bibinfo  {publisher} {Springer},\ \bibinfo {year} {2016})\ \Eprint
  {http://arxiv.org/abs/1701.01554} {arXiv:1701.01554 [hep-ph]} \BibitemShut
  {NoStop}%
\bibitem [{\citenamefont {Weldon}(1982)}]{Weldon:1982aq}%
  \BibitemOpen
  \bibfield  {author} {\bibinfo {author} {\bibfnamefont {H.~Arthur}\
  \bibnamefont {Weldon}},\ }\bibfield  {title} {\enquote {\bibinfo {title}
  {{Covariant Calculations at Finite Temperature: The Relativistic Plasma}},}\
  }\href {\doibase 10.1103/PhysRevD.26.1394} {\bibfield  {journal} {\bibinfo
  {journal} {Phys. Rev. D}\ }\textbf {\bibinfo {volume} {26}},\ \bibinfo
  {pages} {1394} (\bibinfo {year} {1982})}\BibitemShut {NoStop}%
\bibitem [{\citenamefont {Meyer}(2011)}]{Meyer:2011gj}%
  \BibitemOpen
  \bibfield  {author} {\bibinfo {author} {\bibfnamefont {Harvey~B.}\
  \bibnamefont {Meyer}},\ }\bibfield  {title} {\enquote {\bibinfo {title}
  {{Transport Properties of the Quark-Gluon Plasma: A Lattice QCD
  Perspective}},}\ }\href {\doibase 10.1140/epja/i2011-11086-3} {\bibfield
  {journal} {\bibinfo  {journal} {Eur. Phys. J. A}\ }\textbf {\bibinfo {volume}
  {47}},\ \bibinfo {pages} {86} (\bibinfo {year} {2011})},\ \Eprint
  {http://arxiv.org/abs/1104.3708} {arXiv:1104.3708 [hep-lat]} \BibitemShut
  {NoStop}%
\bibitem [{\citenamefont {Ghiglieri}\ \emph {et~al.}(2016)\citenamefont
  {Ghiglieri}, \citenamefont {Kaczmarek}, \citenamefont {Laine},\ and\
  \citenamefont {Meyer}}]{Ghiglieri:2016tvj}%
  \BibitemOpen
  \bibfield  {author} {\bibinfo {author} {\bibfnamefont {J.}~\bibnamefont
  {Ghiglieri}}, \bibinfo {author} {\bibfnamefont {O.}~\bibnamefont
  {Kaczmarek}}, \bibinfo {author} {\bibfnamefont {M.}~\bibnamefont {Laine}}, \
  and\ \bibinfo {author} {\bibfnamefont {F.}~\bibnamefont {Meyer}},\ }\bibfield
   {title} {\enquote {\bibinfo {title} {{Lattice constraints on the thermal
  photon rate}},}\ }\href {\doibase 10.1103/PhysRevD.94.016005} {\bibfield
  {journal} {\bibinfo  {journal} {Phys. Rev. D}\ }\textbf {\bibinfo {volume}
  {94}},\ \bibinfo {pages} {016005} (\bibinfo {year} {2016})},\ \Eprint
  {http://arxiv.org/abs/1604.07544} {arXiv:1604.07544 [hep-lat]} \BibitemShut
  {NoStop}%
\bibitem [{\citenamefont {Brandt}\ \emph {et~al.}(2018)\citenamefont {Brandt},
  \citenamefont {Francis}, \citenamefont {Harris}, \citenamefont {Meyer},\ and\
  \citenamefont {Steinberg}}]{Brandt:2017vgl}%
  \BibitemOpen
  \bibfield  {author} {\bibinfo {author} {\bibfnamefont {Bastian~B.}\
  \bibnamefont {Brandt}}, \bibinfo {author} {\bibfnamefont {Anthony}\
  \bibnamefont {Francis}}, \bibinfo {author} {\bibfnamefont {Tim}\ \bibnamefont
  {Harris}}, \bibinfo {author} {\bibfnamefont {Harvey~B.}\ \bibnamefont
  {Meyer}}, \ and\ \bibinfo {author} {\bibfnamefont {Aman}\ \bibnamefont
  {Steinberg}},\ }\bibfield  {title} {\enquote {\bibinfo {title} {{An estimate
  for the thermal photon rate from lattice QCD}},}\ }\href {\doibase
  10.1051/epjconf/201817507044} {\bibfield  {journal} {\bibinfo  {journal} {EPJ
  Web Conf.}\ }\textbf {\bibinfo {volume} {175}},\ \bibinfo {pages} {07044}
  (\bibinfo {year} {2018})},\ \Eprint {http://arxiv.org/abs/1710.07050}
  {arXiv:1710.07050 [hep-lat]} \BibitemShut {NoStop}%
\bibitem [{\citenamefont {C\`e}\ \emph {et~al.}(2022)\citenamefont {C\`e},
  \citenamefont {Harris}, \citenamefont {Krasniqi}, \citenamefont {Meyer},\
  and\ \citenamefont {T\"or\"ok}}]{Ce:2022fot}%
  \BibitemOpen
  \bibfield  {author} {\bibinfo {author} {\bibfnamefont {Marco}\ \bibnamefont
  {C\`e}}, \bibinfo {author} {\bibfnamefont {Tim}\ \bibnamefont {Harris}},
  \bibinfo {author} {\bibfnamefont {Ardit}\ \bibnamefont {Krasniqi}}, \bibinfo
  {author} {\bibfnamefont {Harvey~B.}\ \bibnamefont {Meyer}}, \ and\ \bibinfo
  {author} {\bibfnamefont {Csaba}\ \bibnamefont {T\"or\"ok}},\ }\bibfield
  {title} {\enquote {\bibinfo {title} {{Photon emissivity of the quark-gluon
  plasma: A lattice QCD analysis of the transverse channel}},}\ }\href
  {\doibase 10.1103/PhysRevD.106.054501} {\bibfield  {journal} {\bibinfo
  {journal} {Phys. Rev. D}\ }\textbf {\bibinfo {volume} {106}},\ \bibinfo
  {pages} {054501} (\bibinfo {year} {2022})},\ \Eprint
  {http://arxiv.org/abs/2205.02821} {arXiv:2205.02821 [hep-lat]} \BibitemShut
  {NoStop}%
\bibitem [{\citenamefont {Meyer}\ \emph {et~al.}(2023)\citenamefont {Meyer},
  \citenamefont {C\`e}, \citenamefont {Harris}, \citenamefont {Krasniqia},\
  and\ \citenamefont {T\"or\"ok}}]{Meyer:2023ntn}%
  \BibitemOpen
  \bibfield  {author} {\bibinfo {author} {\bibfnamefont {Harvey~B.}\
  \bibnamefont {Meyer}}, \bibinfo {author} {\bibfnamefont {Marco}\ \bibnamefont
  {C\`e}}, \bibinfo {author} {\bibfnamefont {Tim}\ \bibnamefont {Harris}},
  \bibinfo {author} {\bibfnamefont {Ardit}\ \bibnamefont {Krasniqia}}, \ and\
  \bibinfo {author} {\bibfnamefont {Csaba}\ \bibnamefont {T\"or\"ok}},\
  }\bibfield  {title} {\enquote {\bibinfo {title} {{Photon and dilepton
  production rate in the quark-gluon plasma from lattice QCD}},}\ }\href
  {\doibase 10.22323/1.430.0186} {\bibfield  {journal} {\bibinfo  {journal}
  {PoS}\ }\textbf {\bibinfo {volume} {LATTICE2022}},\ \bibinfo {pages} {186}
  (\bibinfo {year} {2023})}\BibitemShut {NoStop}%
\bibitem [{\citenamefont {Ali}\ \emph {et~al.}(2024)\citenamefont {Ali},
  \citenamefont {Bala}, \citenamefont {Francis}, \citenamefont {Jackson},
  \citenamefont {Kaczmarek}, \citenamefont {Turnwald}, \citenamefont {Ueding},\
  and\ \citenamefont {Wink}}]{Ali:2024xae}%
  \BibitemOpen
  \bibfield  {author} {\bibinfo {author} {\bibfnamefont {Sajid}\ \bibnamefont
  {Ali}}, \bibinfo {author} {\bibfnamefont {Dibyendu}\ \bibnamefont {Bala}},
  \bibinfo {author} {\bibfnamefont {Anthony}\ \bibnamefont {Francis}}, \bibinfo
  {author} {\bibfnamefont {Greg}\ \bibnamefont {Jackson}}, \bibinfo {author}
  {\bibfnamefont {Olaf}\ \bibnamefont {Kaczmarek}}, \bibinfo {author}
  {\bibfnamefont {Jonas}\ \bibnamefont {Turnwald}}, \bibinfo {author}
  {\bibfnamefont {Tristan}\ \bibnamefont {Ueding}}, \ and\ \bibinfo {author}
  {\bibfnamefont {Nicolas}\ \bibnamefont {Wink}} (\bibinfo {collaboration}
  {HotQCD}),\ }\bibfield  {title} {\enquote {\bibinfo {title} {{Lattice QCD
  estimates of thermal photon production from the QGP}},}\ }\href {\doibase
  10.1103/PhysRevD.110.054518} {\bibfield  {journal} {\bibinfo  {journal}
  {Phys. Rev. D}\ }\textbf {\bibinfo {volume} {110}},\ \bibinfo {pages}
  {054518} (\bibinfo {year} {2024})},\ \Eprint
  {http://arxiv.org/abs/2403.11647} {arXiv:2403.11647 [hep-lat]} \BibitemShut
  {NoStop}%
\bibitem [{\citenamefont {Bazavov}\ \emph {et~al.}(2014)\citenamefont {Bazavov}
  \emph {et~al.}}]{HotQCD:2014kol}%
  \BibitemOpen
  \bibfield  {author} {\bibinfo {author} {\bibfnamefont {A.}~\bibnamefont
  {Bazavov}} \emph {et~al.} (\bibinfo {collaboration} {HotQCD}),\ }\bibfield
  {title} {\enquote {\bibinfo {title} {{Equation of state in ( 2+1 )-flavor
  QCD}},}\ }\href {\doibase 10.1103/PhysRevD.90.094503} {\bibfield  {journal}
  {\bibinfo  {journal} {Phys. Rev. D}\ }\textbf {\bibinfo {volume} {90}},\
  \bibinfo {pages} {094503} (\bibinfo {year} {2014})},\ \Eprint
  {http://arxiv.org/abs/1407.6387} {arXiv:1407.6387 [hep-lat]} \BibitemShut
  {NoStop}%
\bibitem [{Note5()}]{Note5}%
  \BibitemOpen
  \bibinfo {note} {Although these results were at a pion mass $m_\pi \simeq
  320$~MeV and not continuum extrapolated, the lattice was quite large with
  $N_s = 96$ and $a \simeq 7$~GeV$^{-1}$.}\BibitemShut {Stop}%
\bibitem [{\citenamefont {Jackson}(2022)}]{Jackson:2022fqj}%
  \BibitemOpen
  \bibfield  {author} {\bibinfo {author} {\bibfnamefont {Greg}\ \bibnamefont
  {Jackson}},\ }\bibfield  {title} {\enquote {\bibinfo {title} {{Shedding light
  on thermal photon and dilepton production}},}\ }\href {\doibase
  10.1051/epjconf/202227405014} {\bibfield  {journal} {\bibinfo  {journal} {EPJ
  Web Conf.}\ }\textbf {\bibinfo {volume} {274}},\ \bibinfo {pages} {05014}
  (\bibinfo {year} {2022})},\ \Eprint {http://arxiv.org/abs/2211.09575}
  {arXiv:2211.09575 [hep-ph]} \BibitemShut {NoStop}%
\bibitem [{\citenamefont {Faccioli}\ \emph {et~al.}(2010)\citenamefont
  {Faccioli}, \citenamefont {Lourenco}, \citenamefont {Seixas},\ and\
  \citenamefont {Wohri}}]{Faccioli:2010kd}%
  \BibitemOpen
  \bibfield  {author} {\bibinfo {author} {\bibfnamefont {Pietro}\ \bibnamefont
  {Faccioli}}, \bibinfo {author} {\bibfnamefont {Carlos}\ \bibnamefont
  {Lourenco}}, \bibinfo {author} {\bibfnamefont {Joao}\ \bibnamefont {Seixas}},
  \ and\ \bibinfo {author} {\bibfnamefont {Hermine~K.}\ \bibnamefont {Wohri}},\
  }\bibfield  {title} {\enquote {\bibinfo {title} {{Towards the experimental
  clarification of quarkonium polarization}},}\ }\href {\doibase
  10.1140/epjc/s10052-010-1420-5} {\bibfield  {journal} {\bibinfo  {journal}
  {Eur. Phys. J. C}\ }\textbf {\bibinfo {volume} {69}},\ \bibinfo {pages}
  {657--673} (\bibinfo {year} {2010})},\ \Eprint
  {http://arxiv.org/abs/1006.2738} {arXiv:1006.2738 [hep-ph]} \BibitemShut
  {NoStop}%
\bibitem [{\citenamefont {Caron-Huot}(2009)}]{Caron-Huot:2009ypo}%
  \BibitemOpen
  \bibfield  {author} {\bibinfo {author} {\bibfnamefont {S.}~\bibnamefont
  {Caron-Huot}},\ }\bibfield  {title} {\enquote {\bibinfo {title} {{Asymptotics
  of thermal spectral functions}},}\ }\href {\doibase
  10.1103/PhysRevD.79.125009} {\bibfield  {journal} {\bibinfo  {journal} {Phys.
  Rev. D}\ }\textbf {\bibinfo {volume} {79}},\ \bibinfo {pages} {125009}
  (\bibinfo {year} {2009})},\ \Eprint {http://arxiv.org/abs/0903.3958}
  {arXiv:0903.3958 [hep-ph]} \BibitemShut {NoStop}%
\bibitem [{\citenamefont {{Laine, M. and Veps\"al\"ainen, M. and Vuorinen,
  A.}}(2010)}]{Laine:2010tc}%
  \BibitemOpen
  \bibfield  {author} {\bibinfo {author} {\bibnamefont {{Laine, M. and
  Veps\"al\"ainen, M. and Vuorinen, A.}}},\ }\bibfield  {title} {\enquote
  {\bibinfo {title} {{Ultraviolet asymptotics of scalar and pseudoscalar
  correlators in hot Yang-Mills theory}},}\ }\href {\doibase
  10.1007/JHEP10(2010)010} {\bibfield  {journal} {\bibinfo  {journal} {JHEP}\
  }\textbf {\bibinfo {volume} {10}},\ \bibinfo {pages} {010} (\bibinfo {year}
  {2010})},\ \Eprint {http://arxiv.org/abs/1008.3263} {arXiv:1008.3263
  [hep-ph]} \BibitemShut {NoStop}%
\bibitem [{Note6()}]{Note6}%
  \BibitemOpen
  \bibinfo {note} {Note that calculations to NNLO and beyond need to respect
  the fact that $\rho _{\protect \rm L}$ = 0 for $M \to 0$, and $\rho
  _{\protect \rm T} = \rho _{\protect \rm L}$ for $M \to \infty $.}\BibitemShut
  {Stop}%
\bibitem [{\citenamefont {Arslandok}\ \emph {et~al.}(2023)\citenamefont
  {Arslandok} \emph {et~al.}}]{Arslandok:2023utm}%
  \BibitemOpen
  \bibfield  {author} {\bibinfo {author} {\bibfnamefont {M.}~\bibnamefont
  {Arslandok}} \emph {et~al.},\ }\bibfield  {title} {\enquote {\bibinfo {title}
  {{Hot QCD White Paper}},}\ }\href@noop {} {\  (\bibinfo {year} {2023})},\
  \Eprint {http://arxiv.org/abs/2303.17254} {arXiv:2303.17254 [nucl-ex]}
  \BibitemShut {NoStop}%
\bibitem [{\citenamefont {{Schenke, Bj\"orn and Jeon, Sangyong and Gale,
  Charles}}(2010)}]{Schenke:2010nt}%
  \BibitemOpen
  \bibfield  {author} {\bibinfo {author} {\bibnamefont {{Schenke, Bj\"orn and
  Jeon, Sangyong and Gale, Charles}}},\ }\bibfield  {title} {\enquote {\bibinfo
  {title} {{(3+1)D hydrodynamic simulation of relativistic heavy-ion
  collisions}},}\ }\href {\doibase 10.1103/PhysRevC.82.014903} {\bibfield
  {journal} {\bibinfo  {journal} {Phys. Rev.}\ }\textbf {\bibinfo {volume}
  {C82}},\ \bibinfo {pages} {014903} (\bibinfo {year} {2010})},\ \Eprint
  {http://arxiv.org/abs/1004.1408} {arXiv:1004.1408 [hep-ph]} \BibitemShut
  {NoStop}%
\bibitem [{\citenamefont {Schenke}\ \emph {et~al.}(2011)\citenamefont
  {Schenke}, \citenamefont {Jeon},\ and\ \citenamefont
  {Gale}}]{Schenke:2010rr}%
  \BibitemOpen
  \bibfield  {author} {\bibinfo {author} {\bibfnamefont {Bj{\"o}rn}\
  \bibnamefont {Schenke}}, \bibinfo {author} {\bibfnamefont {Sangyong}\
  \bibnamefont {Jeon}}, \ and\ \bibinfo {author} {\bibfnamefont {Charles}\
  \bibnamefont {Gale}},\ }\bibfield  {title} {\enquote {\bibinfo {title}
  {{Elliptic and triangular flow in event-by-event (3+1)D viscous
  hydrodynamics}},}\ }\href {\doibase 10.1103/PhysRevLett.106.042301}
  {\bibfield  {journal} {\bibinfo  {journal} {Phys. Rev. Lett.}\ }\textbf
  {\bibinfo {volume} {106}},\ \bibinfo {pages} {042301} (\bibinfo {year}
  {2011})},\ \Eprint {http://arxiv.org/abs/1009.3244} {arXiv:1009.3244
  [hep-ph]} \BibitemShut {NoStop}%
\bibitem [{\citenamefont {van Hees}\ and\ \citenamefont
  {Rapp}(2008{\natexlab{b}})}]{vanHees:2007ma}%
  \BibitemOpen
  \bibfield  {author} {\bibinfo {author} {\bibfnamefont {H.}~\bibnamefont {van
  Hees}}\ and\ \bibinfo {author} {\bibfnamefont {R.}~\bibnamefont {Rapp}},\
  }\bibfield  {title} {\enquote {\bibinfo {title} {{Thermal Dileptons at
  LHC}},}\ }\href@noop {} {\bibfield  {journal} {\bibinfo  {journal} {J. Phys.
  G}\ }\textbf {\bibinfo {volume} {35}},\ \bibinfo {pages} {054001.153}
  (\bibinfo {year} {2008}{\natexlab{b}})},\ \Eprint
  {http://arxiv.org/abs/0706.4443} {arXiv:0706.4443 [hep-ph]} \BibitemShut
  {NoStop}%
\bibitem [{\citenamefont {{Vujanovic, Gojko and Young, Clint and Schenke,
  Bj\"orn and Rapp, Ralf and Jeon, Sangyong and Gale,
  Charles}}(2014)}]{Vujanovic:2013jpa}%
  \BibitemOpen
  \bibfield  {author} {\bibinfo {author} {\bibnamefont {{Vujanovic, Gojko and
  Young, Clint and Schenke, Bj\"orn and Rapp, Ralf and Jeon, Sangyong and Gale,
  Charles}}},\ }\bibfield  {title} {\enquote {\bibinfo {title} {{Dilepton
  emission in high-energy heavy-ion collisions with viscous hydrodynamics}},}\
  }\href {\doibase 10.1103/PhysRevC.89.034904} {\bibfield  {journal} {\bibinfo
  {journal} {Phys. Rev. C}\ }\textbf {\bibinfo {volume} {89}},\ \bibinfo
  {pages} {034904} (\bibinfo {year} {2014})},\ \Eprint
  {http://arxiv.org/abs/1312.0676} {arXiv:1312.0676 [nucl-th]} \BibitemShut
  {NoStop}%
\end{thebibliography}%

\end{document}